# Estimating Magnitude Completeness in Earthquake Catalogs: A Comparative Study of Catalog-based Methods

Xinyi Wang[1,2], Jiawei Li[2], Ao Feng [1], and Didier Sornette[2]

[1]School of Computer Science, Chengdu University of Information Technology (CUIT), Chengdu, China.

[2]Institute of Risk Analysis, Prediction and Management (Risks-X), Academy for Advanced Interdisciplinary Studies, Southern University of Science and Technology (SUSTech), Shenzhen, China.

Corresponding author: Jiawei Li (lijw@cea-igp.ac.cn); Didier Sornette (didier@sustech.edu.cn)

**Key Points:**

- Evaluated nine catalog-based methods for estimating the completeness magnitude ($M_c$) in both simulated and observed earthquake catalogs.

- Identified MBS-WW as the most reliable method under the assumption of a sharp cut-off $M_c$.

- Introduced BSReLU as an augmented Gutenberg-Richter model that provides a novel probabilistic framework for modeling $M_c$.

## Abstract

Without rigorous attention to the completeness of earthquake catalogs, claims of new discoveries or forecasting skills cannot be deemed credible. Therefore, estimating the completeness magnitude ($M_c$) is a critical step. Among various approaches, catalog-based methods are the simplest, most straightforward, and most commonly used. These methods also often serve as the foundation for more advanced techniques. However, current evaluation frameworks for earthquake forecasts lack a unified simulation strategy for generating catalogs that are independent of specific $M_c$ estimation methods. An effective strategy should also be capable of simulating datasets with non-uniform $M_c$ distributions across both spatial and temporal dimensions. In this study, we assess nine catalog-based methods, including two newly developed approaches, under a robust evaluation framework specifically tailored for this purpose. These methods are tested on datasets with homogeneous and heterogeneous $M_c$ distributions, as well as on real-world earthquake catalogs from China, California, and New Zealand. Among all the methods, the method of *b*-value stability by Woessner and Wiemer (2005), referred to as MBS-WW in this study, demonstrates the best overall performance. For homogeneous $M_c$ datasets, it is the only method capable of providing reliable $M_c$ estimates with a minimum requirement of just 300 events in the complete portion of the catalog. For heterogeneous $M_c$ datasets, MBS-WW produces $M_c$ estimates close to the mean value $M_c$ (3.0) over all grids, provided that sample sizes and grid resolutions are adequate. The performance of MBS-WW is further validated through analysis of observed earthquake catalogs, with mean $M_c$ values aligning closely with the expected $M_c \approx 1.8$. However, these estimates exhibit relatively large uncertainty. The prior model generated by MBS-WW is used as the foundation for generating an updated $M_c$ map with the Bayesian Magnitude of Completeness (BMC) method. This approach reduces uncertainties in $M_c$ estimates from 0.47 to 0.43, enabling the creation of a more reliable $M_c$ map for China. We also introduce, BSReLU, an augmented Gutenberg-Richter model with a novel probabilistic framework for modeling. Unlike traditional methods that rely on a strict cut-off $M_c$, the BSReLU model replaces deterministic estimates of completeness magnitude ($M_c$) with a probabilistic framework that models the smooth transition in detection likelihood from zero to one as earthquake magnitudes increase. This approach effectively captures the gradual transition between incomplete and complete data, overcoming the limitations of cut-off models. Continued exploration of BSReLU's potential is recommended for future research. By evaluating the limitations of these foundational catalog-based methods in estimating $M_c$, this study seeks to refine our understanding of their appropriate applications, offering a clearer, unbiased perspective on seismicity patterns, earthquake forecasting, and hazard mitigation through improved observational data quality.

## Plain Language Summary

In earthquake studies, determining the smallest magnitude at which all earthquakes are detected, known as the completeness magnitude ($M_c$), is essential for accurate analysis. Various methods exist to estimate $M_c$, but it is crucial to identify which are most reliable under different conditions. This study evaluates nine catalog-based methods, including two newly developed approaches, by testing them on both simulated and real-world earthquake data from regions like China, California, and New Zealand. Among these, the method of b-value stability by Woessner and Wiemer (2005), referred to as MBS-WW, demonstrated superior performance. For homogeneous $M_c$ datasets, it is the only method capable of providing reliable $M_c$ estimates with a minimum requirement of just 300 events in the complete portion of the catalog. For heterogeneous $M_c$ datasets, MBS-WW produces $M_c$ estimates close to the mean value $M_c$ (3.0) over all grids, provided that sample sizes and grid resolutions are adequate. We introduce, BSReLU, a novel approach that models the transition from complete to incomplete in the catalogue as a gradual (rather than a sharp) transition. This probabilistic framework is better suited to represent the engineering and physical steps of earthquake detection. Understanding better the strengths and limitations of the discussed methods is likely to improve earthquake forecasting and hazard assessments.

**1 Introduction**

The completeness magnitude ($M_c$) is the threshold magnitude above which all earthquakes are reliably recorded within a specified spatial-temporal window (Rydelek and Sacks, 1989; Mignan and Woessner, 2012). It represents the smallest magnitude at which the catalog is deemed "complete," meaning all earthquakes of this magnitude or larger are recorded with near certainty (probability close to 1). These events adhere to the Gutenberg-Richter (GR) law, which describes the dependence of the complementary cumulative frequency-magnitude distribution (CCFMD) as being a decaying exponential function of the magnitude:

$$\text{CCFMD}(M) = 10^{a-bM}, \ M \geq M_c \qquad (1)$$

Here, $a$ and $b$ are constants, with $b$, the slope of the GR law, typically close to 1. Below the magnitude of completeness, smaller earthquakes are recorded with a probability smaller than 1, depending on the detection capabilities of the seismic monitoring network.

Accurately estimating $M_c$ is crucial for reliable seismic hazard assessments, earthquake forecasting, and understanding the underlying processes that govern seismicity. Indeed, many—if not most—claims of discovering new patterns, structures, effects, or laws in earthquake forecasting have ultimately been traced back to biases stemming from incompleteness effects in earthquake catalogs. Without rigorous attention to the completeness of these catalogs, no claim of new discoveries or forecasting skills—whether derived from back tests, statistical analyses, or advanced modeling—can be deemed credible. The integrity of earthquake catalog data is fundamental, as any overlooked gaps or inconsistencies may lead to false positives, misinterpretations, and unreliable predictions. Only by addressing these issues with the utmost diligence can meaningful advancements in seismic research and forecasting be achieved. Therefore, an accurate estimation of $M_c$ has become one of the essential first steps in nearly all robust studies of earthquake statistical physics (Mignan and Woessner, 2012).

Many methods have been developed to assess $M_c$, which can be broadly grouped into three categories:

(a) Catalog-based methods rely solely on earthquake catalogs as input data. These methods are typically easier to implement and use data that are more readily available. Common catalog-based methods include the maximum curvature (MAXC; Wyss et al., 1999) method, goodness-of-fit (GFT) tests (Wiemer and Wyss, 2000), the b-value stability method (MBS; Cao and Gao, 2002; Woessner and Wiemer, 2005), the entire magnitude range (EMR) method (Woessner and Wiemer, 2005), median-based analysis of the segment slope (MBASS; Amorèse, 2007), and day-to-night noise modulation (Rydelek and Sacks , 1989). Among them, MAXC and MBASS are non-parametric techniques based on the frequency-magnitude distribution (FMD). In recent years, Herrmann and Marzocchi (2021) proposed a method for analyzing $M_c$ by testing the exponentiality of FMD, using the Lilliefors test (Lilliefors, 1969). Building on this approach,

Taroni (2023) explored whether the method for estimating $M_c$ by testing the exponentiality could be applied to earthquake catalogs composed of mixed exponential distributions, such as those containing events with varying *b*-values. However, much like the Kolmogorov-Smirnov statistic-based approach proposed by Clauset et al. (2009) for determining the minimum cut-off sample size necessary to preserve a power law distribution, these test-based methods are computationally intensive. This is due to the requirement of generating a large number of synthetic samples to compute *p*-values. Consequently, their computational efficiency is notably low.

(b) Network-based methods, such as the Probabilistic Magnitude of Completeness (PMC; Schorlemmer and Woessner, 2008) and others discussed in Section 5.2 of Mignan and Woessner (2012), approach $M_c$ from a network detection perspective. The primary goal of these methods is not to define the completeness level of the catalog, but rather to estimate the probability that an earthquake can be detected. These methods require input data such as seismic waveforms, phase-pick data, station metadata, and attenuation relations used for magnitude determination. However, they face challenges related to both data availability and the complexity of computation and result interpretation.

(c) Hybrid methods combine one or more catalog-based methods with other techniques. For example, Mignan and Woessner (2012) proposed a combined approach to estimating $M_c$: when the GFT test does not reach the 95% confidence level, the fitting level is lowered to 90%. If $M_c$ cannot be estimated with this adjustment, the MAXC method is used. This strategy is implemented in the ZMAP software (Wiemer, 2001). Another example is the Bayesian Magnitude of Completeness (BMC) proposed by Mignan et al. (2011), which estimates posterior $M_c$ by integrating $M_c$ estimates obtained using one of the catalog-based methods (e.g., MAXC in Mignan et al., 2011; Feng et al., 2022; MBASS in Li et al., 2024) with prior Bayesian information, weighted by their respective standard deviations. This prior information is derived from the relationship between the density of the seismic network and $M_c$ estimates. Hence, hybrid methods strongly rely on the reliability and robustness of the catalog-based methods used as their foundational algorithms.

In summary, catalog-based methods are straightforward to implement using readily available data and often serve as the foundation for more advanced $M_c$ analysis techniques. Therefore, it is crucial to understand which of these methods are more reliable, and under what circumstances. Existing comparisons primarily assess the accuracy and stability of these methods using observed seismicity with unkown $M_c$ or simulated seismicity with known $M_c$ (Woessner and Wiemer, 2005; Mignan and Woessner, 2012; Herrmann and Marzocchi, 2021; Taroni, 2023; Lombardi, 2021; Pavlenko and Zavyalov, 2022). For observed data, the consistency of $M_c$ estimates across different methods applied to the same data is also tested (Kagan, 2003; Husen and Hardebeck, 2010; Herrmann and Marzocchi, 2021; Mignan and Woessner, 2012). However, most of these comparison strategies overlook the critical fact that $M_c$ can vary over time and space due to heterogeneities in station distribution, site conditions, and sensor sensitivity (e.g., Mignan et al., 2011; Mignan et al., 2012; Godano et al., 2021; Feng et al., 2022; Li et al., 2024). Furthermore, existing models commonly used to synthesize incomplete data assume that FMD is a mixture of Normal (Gaussian) and power law (Gutenberg-Richter) components, with the magnitude of transition being identified as $M_c$. However, this synthesis model represents a specification of the EMR method (Section 2), and thus, a mis-specification of other methods. Additionally, the asymmetry in incomplete segments often leads to discontinuities at the transformed magnitude $M_c$ between the Gaussian and power-law components. Additionally, other approaches have been proposed to model the incomplete portion of the FMD, such as using another exponential distribution (ANgular model; Mignan, 2012) or a quadratic polynomial distribution (POL model; García-Hernández et al., 2019). However, these models differ significantly from the actual seismicity distributions we observe (e.g., as shown in Figure S2). Therefore, there is a need for a unified simulation strategy for generating data that is not tied to any specific $M_c$ estimation method, and this strategy should also simulate data with a non-uniform $M_c$ distribution.

The purpose of the present paper is to assess the accuracy and reliability of commonly used catalog-based methods for estimating $M_c$ using both simulated and observed data that reflect the heterogeneous temporal and spatial distribution of $M_c$. We introduce two new catalog-based methods for estimating $M_c$, drawing on our expertise in both statistical testing and computer science, and compare them with other existing methods. Additionally, we propose a novel seismicity pruning framework that not only simulates incomplete seismicity through a pruning function but also incorporates the modeling of the spatiotemporal distribution of this incompleteness. By examining the limitations of these foundational catalog-based methods in estimating $M_c$, we aim to enhance our understanding of how these methods should be applied, ultimately providing an unbiased perspective on seismicity patterns, earthquake predictability, and earthquake hazard mitigation based on accurate knowledge of observational data quality.

## 2 Methods for estimating completeness magnitude using catalog

In this section, we appraise nine catalog-based methods, including two new methods developed by drawing on our expertise in both statistical testing and computer science. The present paper does not discuss the day-to-night noise modulation method proposed by Rydelek and Sacks (1989), although it is also a catalog-based method. This is because the method requires the prior removal of other non-random features from the earthquake catalog, such as earthquake clusters. These requirements impose significant limitations on its applicability (Wiemer and Wyss, 2003).

(1) MAXC

The maximum curvature (MAXC) method, proposed by Wyss et al. (1999), determines $M_c$ by identifying the magnitude corresponding to the point of maximum curvature, calculated as the maximum value of the first derivative of the FMD, which is itself the first derivative of the CCFMD, making it a fast and straightforward approach. Generally, the $M_c$ obtained by MAXC, which is typically used as the initial condition in many of the algorithms introduced later, corresponds to the magnitude with the highest event frequency in the FMD.

(2) GFT-95%

The goodness-of-fit tests (GFT), proposed by Wiemer and Wyss (2000), estimate $M_c$ by assessing the goodness of fit between the observed and predicted cumulative frequency-magnitude distributions, namely $CFMD^{obs}(M)$ and $CFMD^{pred}(M)$, by systematically scanning across magnitudes. These magnitudes are used as the cut-off magnitude, $M_{co}$, to calculate $GFT(M_{co})$ as follows:

$$\text{GFT}(M_{\text{co}}) = 100 - \left[ \frac{\sum\limits_{M=M_{\text{co}}}^{M_{\max}} |\text{CFMD}^{\text{obs}}(M) - \text{CFMD}^{\text{pred}}(M)|}{\sum\limits_{M=M_{\text{co}}}^{M_{\max}} \text{CFMD}^{\text{obs}}(M)} 100 \right] \quad (2)$$

Here, $M_{\max}$ is the maximum magnitude of the catalog events. The GFT-95% defines $M_c$ as the $M_{\text{co}}$ at which GFT($M_{\text{co}}$) first falls below the 95% fitting level (or confidence level) as magnitudes are scanned by decreasing values.

(3) GFT-90%

GFT-90% is a variant of the GFT test, defining $M_c$ as the $M_{\text{co}}$ where GFT($M_{\text{co}}$) drops below the 90% fitting level (or confidence level). It typically yields a smaller $M_c$ estimate than GFT-95%.

(4) KST-95%

The Kolmogorov-Smirnov statistic is commonly employed to measure the degree of divergence between a sample dataset and its corresponding theoretical distribution (Gibbons and Chakraborty, 2003; Clauset et al., 2009; Arshad et al, 2010; Corral et al., 2011; Molnár and Szokol, 2014; Baumgartner and Kolassa, 2021; Lombardi, 2021). In the present paper, we modify the numerator of the GFT by replacing it with the Kolmogorov-Smirnov statistic. The denominator is defined as the value of CFMD$^{\text{obs}}$ at the magnitude $M_{\max(\text{KS})}$ of maximum distance between CFMD$^{\text{obs}}(M)$ and CFMD$^{\text{pred}}(M)$. Building upon this, we define a Kolmogorov-Smirnov statistic test (KST) as a function of $M_{\text{co}}$, KST($M_{\text{co}}$), as follows:

$$\text{KST}(M_{\text{co}}) = 100 - \left[ \frac{|\text{CFMD}^{\text{obs}}(M_{\max(\text{KS})}) - \text{CFMD}^{\text{pred}}(M_{\max(\text{KS})})|}{\text{CFMD}^{\text{obs}}(M_{\max(\text{KS})})} 100 \right] \quad (3)$$

KST-95% defines $M_c$ as the $M_{\text{co}}$ where KST($M_{\text{co}}$) first drops below the 95% confidence level, when scanning magnitudes by decreasing values.

(5) MBS-CG

Cao and Gao (2002) proposed a method for estimating $M_c$ by analyzing the stability of the $b$-value as a function of $M_{\text{co}}$. This method, referred to as the method of $b$-value stability (MBS) by Woessner and Wiemer (2005), is investigated in the present paper. To distinguish it from an improved version of MBS introduced later, we refer to it here as MBS-CG. The theoretical foundation of this method is that the estimated $b$-value behaves as follows: it increases with $M_{\text{co}}$ when $M_{\text{co}} < M_c$, stabilizes when $M_{\text{co}} \geq M_c$, and fluctuates significantly when $M_{\text{co}} \gg M_c$. Specifically, for $M_{\text{co}} < M_c$, the FMD deviates from the GR law, falling below the GR law extrapolation at smaller magnitudes, leading to inaccurate $b$-value estimates. As $M_{\text{co}}$ approaches $M_c$, the estimated $b$-value gradually converges to its true value and remains stable when $M_{\text{co}} \geq M_c$. However, as $M_{\text{co}}$ increases significantly beyond $M_c$, the sample size decreases, particularly in the right tail of the distribution, causing strong fluctuations in the $b$-value. Cao and Gao (2002) defined $M_c$ as the magnitude at which the change in $b$-value ($\Delta b$) between two consecutive $M_{\text{co}}$ values first becomes less than 0.03 as $M_{\text{co}}$ is increased.

(6) MBS-WW

Woessner and Wiemer (2005) identified that the instability of the MBS-CG method was due to the significant variability in the event frequency in each magnitude bin. To enhance the reliability of the *b*-value stability estimation method, they incorporated the more objective *b*-value uncertainty index, *δb*, proposed by Shi and Bolt (1982), as a criterion:

$$\delta b = 2.3 b^2 \sqrt{\frac{\sum_{i=1}^{N}(M_i - M_{\text{ave}})^2}{N_{\text{evt}}(N_{\text{evt}} - 1)}} \quad (4)$$

Here, $M_{\text{ave}}$ is the average magnitude of the catalog events, $N_{\text{evt}}$ is the number of events. $M_c$ is then defined as the first $M_{\text{co}}$ at which $\Delta b(M_{\text{co}}) = |b(M_{\text{co}}) - b_{\text{ave}}| \leq \delta b$ as $M_{\text{co}}$ is increased. The arithmetic mean of the *b*-values, $b_{\text{ave}}$, is calculated by averaging all *b*-values within a specified magnitude range ($\Delta M$) according to:

$$b_{\text{ave}} = \frac{\sum_{M=M_{\text{co}}}^{M_{\text{co}}+\Delta M} b(M) dM}{\Delta M} \quad (5)$$

Typically, the magnitude bin *dM* is set to 0.1 and the magnitude range $\Delta M$ is set to 0.5. To distinguish this improved method introduced by Woessner and Wiemer (2005) from the original MBS-CG, we refer to it as MBS-WW.

(7) EMR

The entire magnitude range (EMR) method, proposed by Woessner and Wiemer (2005), estimates $M_c$ by simultaneously considering both the complete and incomplete parts of the FMD, aiming to replicate the properties of the entire distribution. For the complete part ($M \geq M_c$), the GR law distribution of Equation (1) is assumed. For the incomplete part ($M < M_c$), a Normal cumulative distribution function $q(M| \mu, \sigma)$ is employed to model the detection probability of the seismic network for earthquakes of small magnitudes. The function $q(M| \mu, \sigma)$ is expressed as:

$$q(M \mid \mu, \sigma) = \begin{cases} \frac{1}{\sigma\sqrt{2\pi}} \int_{-\infty}^{M_c} \exp\left[-\frac{(M-\mu)^2}{2\sigma^2}\right] dM, & M < M_c \\ 1, & M \geq M_c \end{cases} \quad (6)$$

Here, *μ* represents the magnitude at which 50% of earthquakes are detected by the seismic network, while σ represents the width of the identified incomplete portion of the magnitude range (i.e., the standard deviation of the normal cumulative distribution function). A larger σ implies a slower decline in the monitoring capacity of the seismic network as the magnitude decreases. For $M \geq M_c$, the detection probability is assumed to be 1, aligning with the CCFMD described by Equation (1). This dual approach allows the EMR method to effectively characterize both the complete and incomplete portions, providing a comprehensive estimate of $M_c$.

(8) MBASS

The median-based analysis of the segment slope (MBASS) method, proposed by Amorèse (2007), employs a non-parametric statistical approach to detect change points in the FMD. This method utilizes the multiple change-point procedure developed by Lanzante (1996), an iterative technique for identifying multiple change points in time series data. Lanzante's method identifies points at which the median of a time series changes (Siegel and Castellan, 1988). Amorèse (2007) applied this approach to the segmented slopes of FMDs. The process begins by dividing the set of magnitudes into $n_{seg}$ segments, each defined by magnitude boundaries $M_i$ ($i = 1, ... , n_{seg}$). The slope $s(M_i)$ for the $i$-th segment is then calculated based on the FMD within these boundaries:

$$s(M_i) = \frac{\log[\text{FMD}(M_{i-1})] - \log[\text{FMD}(M_i)]}{M_{i-1} - M_i} \tag{7}$$

Next, the rank sum $SR_i$ for the $i$-th segment is calculated as the sum from 1 up to its position in the ordered set $s\{M_q\}$, where $q = 1, ..., n_{seg}$. This is followed by adjusting the rank sum to obtain $SA_i$:

$$SA_i = |(2\sum_{j=1}^{i} SR_j) - i(n_{seg} + 1)| \tag{8}$$

The maximum value of $SA_i$ is then identified, and the corresponding point $n_{max}(SA)$ is used to divide $s(M)$ into two groups. A null hypothesis is formulated that assumes no discontinuity in $s(M)$ at $n_{max}(SA)$. The Wilcoxon rank-sum test (Wilcoxon, 1945) is applied to determine whether the null hypothesis can be rejected. The test is performed at a significance level of 0.05. If a $p$-value < 0.05 is obtained, the magnitude associated with the minimum $p$-value is identified as the estimated $M_c$. The MBASS technique can identify multiple discontinuities in the FMD. The primary discontinuity typically corresponds to $M_c$, while other discontinuities may indicate high-magnitude breakpoints (Wesnousky, 1994).

(9) BSReLU

We propose an augmented Gutenberg-Richter model to characterize the CCFMD of earthquake catalogs that include incomplete portions, according to:

$$\text{CCFMD}(M) = 10^{a - b\left[M_c + \sigma_{M_c} \text{gReLU}\left(\frac{M - M_c}{\sigma_{M_c}}\right)\right]} \tag{9}$$

where gReLU($x$) is a function that converges to 0 for $x \to -\infty$ and to $x$ for $x \to +\infty$. A special case is ReLU($x$), short for Rectified Linear Unit, which is a commonly used activation function in deep learning, designed to introduce nonlinearity into the model. It is defined such that ReLU($x$) = 0 for $x < 0$ and ReLU($x$) = $x$ for $x \geq 0$. With the choice gReLU($x$) = ReLU($x$), CCFMD($M$) = 10$a$ - $bM$ for $M \geq M_c$ and CCFMD($M$) = 10$a$ - $bM_c$ for $M < M_c$. This corresponds to the exact Gutenberg-Richter law for $M \geq M_c$, with no earthquake considered for $M < M_c$ since CCFMD($M$) is constant for $M < M_c$.

Here, we seek a function that smoothly transitions between the Gutenberg-Richter law for $M \geq M_c$ and a gradual depletion of earthquake occurrences for $M < M_c$. In addition to the condition to converge to 0 for $x \to -\infty$ and to $x$ for $x \to +\infty$, the smoothing function must ensure that CCFMD($M$) is monotonous. Hence, gReLU($x$) must be a monotonously increasing function of $x$. A function that satisfies all these requirement turns out to be a function taken from the formula giving the price of a financial call option according to the Black-Scholes model (Black and Scholes, 1973):

$$\text{BSReLU}(x) = (x + M_c)\Phi\left(\frac{\ln[(x + M_c)/M_c]}{\sigma_{M_c}} + \frac{\sigma_{M_c}}{2}\right) - M_c\Phi\left(\frac{\ln[(x + M_c)/M_c]}{\sigma_{M_c}} - \frac{\sigma_{M_c}}{2}\right) \tag{10}$$

In the limit $\sigma_{M_c} \to 0$, BSReLU($x$) tends to ReLU($x$). Substituting Equation (10) into Equation (9), the augmented Gutenberg-Richter model incorporates not only the $a$-value and $b$-value from the

Gutenberg-Richter law of Equation (1) but also introduces two additional parameters, $M_c$ and $\sigma_{Mc}$, which quantify the completeness magnitude. In this model, $M_c$ is no longer treated as a hard threshold that separates the complete and incomplete parts of the CCFMD. Instead, it is modeled as the mean position of the transition that has a width given by $\sigma_{Mc}$. This replaces the point representation of $M_c$ by a probabilistic description of $M_c$. In other words, the BSReLU model simultaneously captures the Gutenberg-Richter behavior of the logarithm of CCFMD, which asymptotically aligns with a linear trend for $M \geq M_c$, while also integrating a probabilistic representation of the completeness magnitude. This implies that, even for magnitudes larger than $M_c$, there is still a small but nonzero probability of missing some large earthquakes. Conversely, this also indicates that, for magnitudes smaller than $M_c$, the probability of detecting small earthquakes is not zero, but decreases progressively as the magnitude diminishes. By integrating these features, the BSReLU model provides a more physically grounded framework for describing the complexities of real-world earthquake catalogs, particularly those with incomplete entries, offering greater flexibility and comprehensiveness in modeling.

## 3 Earthquake catalog

In this section, we introduce the synthetic simulated earthquake catalogs and the observed earthquake catalogs used in our study.

For the synthetic catalogs, we first generate a complete dataset of earthquakes with their magnitudes drawn from the GR law, assuming that no event is missed. Next, we implement a pruning operation to discard data, simulating an incomplete dataset. Finally, we use this method across multiple grids and combine the data to construct synthetic earthquake catalogs with an heterogeneous $M_c$ distribution.

For the observed earthquake catalogs, we select datasets from four regions in Chinese mainland and two additional regions, California and New Zealand, making a total of six regions.

### 3.1 Simulated earthquake catalog

We generate synthetic earthquake catalog magnitude samples using the normalized magnitude probability density function (or frequency-magnitude distribution, FMD) $P(M)$, which is derived from the complementary cumulative density function (CCDF) given by Equation (1). The normalized FMD $P(M)$ is expressed as:

$$P(M) = \frac{b\ln(10)10^{-bM}}{10^{-bM_{\min}} - 10^{-bM_{\max}}}, M_{\min} \leq M \leq M_{\max} \tag{11}$$

Here, $M_{\min}$ and $M_{\max}$ represent the minimum and maximum magnitudes of earthquakes in the synthetic earthquake catalog. In this study, the parameters for simulating the catalog are set to $b = 1.0$, $M_{\min} = 1.0$, and $M_{\max} = 9.0$. Additionally, the number of simulated earthquake events is set to $10^k$ (where $k$ takes on values $k = 3, 4, 5, 6$) to represent seismicity counts at varying spatial and temporal scales. Examples of FMD's and CCFMD's of simulated catalogs are shown in Panels (I) of Figures S1(a) through 1(d).

After obtaining the FMD of the synthetic catalog with a magnitude bin size of 0.1, we apply the pruning function described by Equation (12) below to simulate the incomplete observation of earthquake catalogs within a given magnitude range ($m_{\min}$, $m_{\max}$). Please note that $m$ here refers to the magnitude range used in the pruning function, which is different from $M$, the earthquake magnitude in the synthetic catalog. To model catalogue incompleteness, we construct a linear magnitude pruning function $y(m)$ defined as:

$$y(m) = c_1 m + c_2 \qquad (12)$$

This function specifies the percentage of missing earthquakes of a given magnitude $m$ ($m_{min} \leq m \leq m_{max}$) relative to the complete dataset. The parameters $c_1$ and $c_2$ are determined by the boundary conditions: at $m = m_{min}$, $y(m) = 1$, and at $m = m_{max}$, $y(m) = 0$. In other words, the function assumes that, at $m = m_{min}$ and lower, 100% of the data is missing, and at $m = m_{max}$ and higher, 100% of the data is recorded. Between these two extremes, the data loss ratio decreases linearly from $m_{min}$ to $m_{max}$. Catalogues of earthquake with a given data loss ratio are generated using random numbers following a Poisson distribution with $y(m)$ as the mean value. In this study, we set $m_{min} = M_{min} = 1.0$, while $m_{max}$ varies from $m_{min} + 0.1$ to 5.0. Here, $m_{max}$ represents the true minimum completeness magnitude $M_c^{true}$ after data pruning.

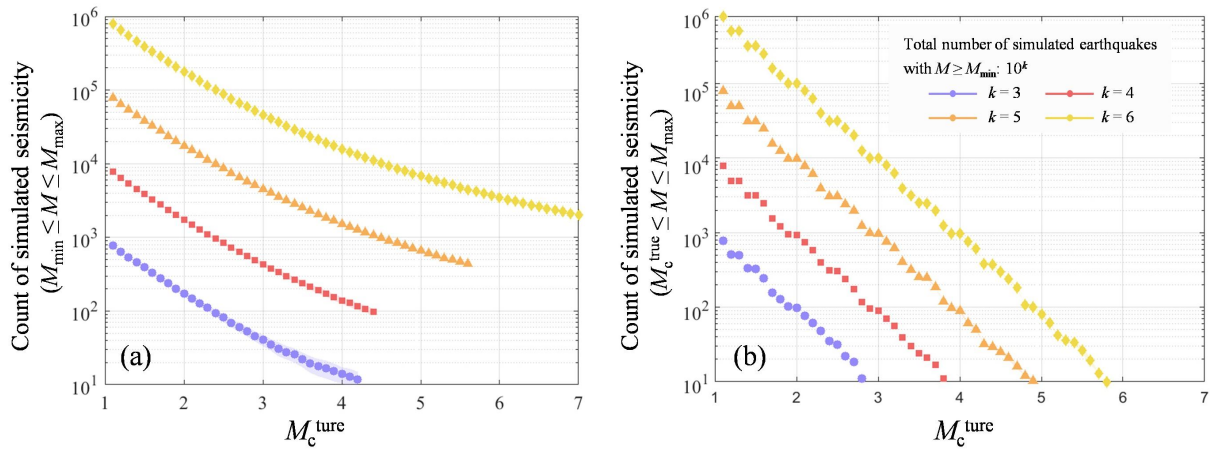

**Figure 1**. Mean (solid symbols) and standard deviation (shaded area) of earthquake counts from 200 sets of simulated earthquake catalogs after data pruning as a function of the true completeness magnitude $M_c^{true}$: (a) for magnitudes between the minimum and maximum simulated magnitudes ($M_{min} = 1.0$ and $M_{max} = 9.0$), and (b) for magnitudes greater than or equal to $M_c^{true}$. As an example shown in Figure S1 for $M_c^{true} = 2.5$, a linear data pruning function is applied iteratively three times between $M_{min}$ and $M_c^{true}$ to simulate data loss, modifying the pure Gutenberg-Richter relationship between $M_{min}$ and $M_{max}$, and mimicking incomplete seismicity records in real observations. The number of lost seismicity is determined by generating random numbers from a Poisson distribution, using the values calculated by the data pruning function as the mean, which results in the variance (shaded area) observed in the figure. The narrowness of the shaded area is primarily due to the logarithmic scale of the vertical axis. The different colors represent the total number of simulated earthquakes with $M \geq M_{min}$, given by $10^k$ with $k = 3, 4, 5, 6$. For details on data pruning, refer to the main text.

Performing multiple data pruning operations is insightful, as it allows us to simulate a range of incomplete catalogs. Panels (II) to (V) of Figures S1(a) through 1(d) illustrate the results of applying the pruning function iteratively from $m_{min}$ to $m_{max}$ for 1 to 4 iterations. Figures 1a and

1b, respectively, show the mean and standard deviation of earthquake counts derived from 200 sets of simulated earthquake catalogs as a function of the $M_c^{true}$ for magnitudes between $M_{min}$ = 1.0 and $M_{max}$ = 9.0, and for magnitudes greater than or equal to $M_c^{true}$ (or $m_{max}$) after 3 iterations of data pruning.

After applying the above data pruning, each dataset is characterized by a uniform (or homogeneous) $M_c^{true}$ (or $m_{max}$). To simulate datasets with spatially and temporally heterogeneous $M_c$ values, the process essentially involves repeating the aforementioned steps for generating $M_c^{true}$ within multiple grids, where each grid is assigned a different $M_c^{true}$. In this study, for each scenario with total data sizes $10^k$ ($k$ = 3, 4, 5, 6), the data are first generated using Equation (11) without pruning and are then evenly distributed across $2^n$ grids ($n$ = 1, 2, ..., 10). Within each grid, the data are pruned, and the $M_c^{true}$ (or $m_{max}$) for each grid is determined as follows: the first and last grids are fixed with $M_c^{true}$ = 1.0 and 5.0, respectively. For the remaining grids, $M_c^{true}$ is linearly interpolated based on the grid index. For example, in the case of $2^2$ = 4 grids, the first grid has $M_c^{true}$ = 1.0, and the fourth grid has $M_c^{true}$ = 5.0. The second and third grids have $M_c^{true}$ = 2.3 and $M_c^{true}$ = 3.7, respectively, determined by linear interpolation. This construction ensures that the mean $M_c^{true}$ across all grids is 3.0, and half of the grids have $M_c^{true} \leq 3.0$, while the other half have $M_c^{true} > 3.0$. Once $M_c^{true}$ values for all grids are determined, Equation (12) is applied to prune the data in each grid iteratively three times. Finally, the pruned data from all grids are combined to create a dataset representing seismicity with heterogeneous $M_c$ values. This approach enables the generation of seismicity that captures spatial and/or temporal variability in $M_c$.

### 3.2 Observed earthquake catalog

We select earthquake catalogs from four regions in the Chinese mainland, as well as California and New Zealand. For the Chinese mainland, we consider the Beijing-Tianjin-Hebei region, the southeastern coastal region, the Sichuan-Yunnan region, and northern Xinjiang. The spatial boundaries of these regions are delineated using the $M_c \approx 1.8$ contour lines derived from the posterior $M_c$ map generated by the BMC method in Li et al. (2024). The data for these regions were sourced from the China Earthquake Networks Center (CENC) and span from January 1, 2009, to November 18, 2024. The number of earthquake events in these catalogs are as follows: 66,380 in the Beijing-Tianjin-Hebei region, 31,421 in the southeastern coastal region, 602,858 in the Sichuan-Yunnan region, and 60,830 in northern Xinjiang. Among these events, the counts of earthquakes with $M \geq 3.0$ are 362, 109, 5,641, and 998, respectively. The California catalog was obtained from the Advanced National Seismic System (ANSS) Comprehensive Earthquake Catalog (ComCat; U.S. Geological Survey, 2017) and spans from January 1, 2009, to November 2, 2023. The spatial boundary is delineated using the $M_c \approx 1.8$ contour lines derived from the posterior $M_c$ map generated by the BMC method in Tormann et al. (2014). This catalog contains 270,458 earthquake events, of which 772 have $M \geq 3$. For New Zealand, the catalog was sourced from the GeoNet Earthquake Catalog (GNS Science, 2022) and spans from January 1, 2009, to December 12, 2023. The spatial boundary is delineated using the station distribution map in Petersen (2011), focusing on areas with dense and evenly distributed seismic stations. This catalog includes 210,549 earthquake events, with 8,450 having $M \geq 3$. Figure S2 shows the complementary cumulative and density distribution functions of their magnitudes for the observed seismicity in the six study regions.

# 4 Results

In this section, we present the performance of nine catalog-based methods using simulated catalogs with both homogeneous and heterogeneous $M_c$ distributions, as well as empirical catalogs.

## 4.1 Estimating $M_c$ for simulated seismicity with homogeneous $M_c$

We found that the CCFMD and FMD of the simulated catalogs, after being iteratively pruned three times, closely resemble those of observed seismicity, as shown in Panels (IV) of Figures S1(a) through 1(d) and compared with Figure S2. Therefore, we primarily report the results based on the dataset iteratively pruned three times in the main text. Figure 2 shows the results of estimating the completeness magnitude $M_c^{pred}$ using nine catalog-based methods as a function of the true completeness magnitude $M_c^{true}$ for the simulated catalogs with homogeneous $M_c$, after being iteratively pruned three times. The solid symbols and shaded areas represent the mean and uncertainties of $M_c^{pred}$ derived from 200 sets of simulated catalogs. Figures S3, S4, and S5 show the results of $M_c^{pred}$ as a function of $M_c^{true}$ for the simulated catalogs with homogeneous $M_c$, after being iteratively pruned once, twice, and four times, respectively.

With the exception of EMR, which lacks $M_c^{pred}$ for $M_c^{true}$ range between 1.0 and 1.7, due to its minimum data requirement for fitting the incomplete portion ($M < M_c^{true}$), all methods provide reasonably accurate $M_c^{pred}$ when $M_c^{true}$ is small ($M_c^{true} \leq 2.0$). However, as $M_c^{true}$ increases, all methods tend to underestimate $M_c$, regardless of the dataset size, except for BSReLU for $k = 4$ and $k = 5$, and EMR for $k = 3$. All eight methods, except for MBS-WW, fail to obtain reliable $M_c^{pred}$ when $M_c^{true}$ is larger than or equal to 3.5. This suggests that these methods may not be able to reliably and accurately estimate the short-term incompleteness of aftershock sequences following major earthquakes.

$M_c^{pred}$ obtained by MAXC begins to deviate from $M_c^{true}$ around $M_c^{true} = 2.0$, regardless of the dataset size. This behavior is due to the method's sensitivity to the peak of the FMD. The results from GFT-90% and GFT-95% show that increasing the confidence level significantly improves the performance. The performance of KST-95% is essentially consistent with that of GFT-95%, meaning that, if an estimation error of 0.2 is acceptable, reliable $M_c^{pred}$ are obtained for $k = 3, 4, 5, 6$ at $M \leq M_c^{true} \approx 1.7, 2.5, 3.3, 3.3$, respectively (corresponding to the minimum number of events in the complete portion in Figure 1b, approximately 200, 300, 400, and 5000). For MBS-CG, its simplistic definition of $b$-value stability leads to the largest deviation of $M_c^{pred}$ from $M_c^{true}$. The obtained $M_c^{pred}$ values are only reliable for $k = 4, 5, 6$ when $M \leq M_c^{true} \approx 2.5$, and for $k = 3$ when $M \leq M_c^{true} \approx 2.0$. The drawbacks of EMR are very apparent, as it requires a sufficient amount of data in the incomplete region to fit the Gaussian distribution. This causes EMR to fail when $M_c^{true}$ is very close to $M_{min}$ and results in unstable estimates for small datasets ($k = 3$). For $k = 4, 5, 6$, EMR typically gives severely underestimated $M_c^{pred}$ values when $M > M_c^{true} \approx 2.5$. MBS-WW appears to be the most promising $M_c$ estimation algorithm, as reliable results are obtained for $k = 3, 4, 5, 6$ at $M \leq M_c^{true} \approx 1.9, 2.7, 3.7, 4.5$, respectively, with the minimum number of events in the complete portion (Figure 1b) being approximately 100, 200, 200, and 300. This indicates that, when the number of events with $M \geq M_c$ in a catalog exceeds approximately 300, MBS-WW can provide very accurate and stable $M_c^{pred}$. MBASS exhibits a similar performance to MBS-WW, but the stability of the estimated $M_c^{pred}$ is notably inferior to that of MBS-WW. BSReLU demonstrates strong stability in $M_c^{pred}$, although $M_c^{pred}$ for $k = 3$ and 6 show some deviation starting around $M_c^{true} \approx 2.0$, for $k = 4$ and 5, deviations begin at $M_c^{true} =$

2.3 and 3.0, respectively. The results for synthetic catalogs obtained by iteratively pruning one, two, and four times show similar trends, with slight variations depending on the degree of data incompleteness controlled by the number of pruning operations.

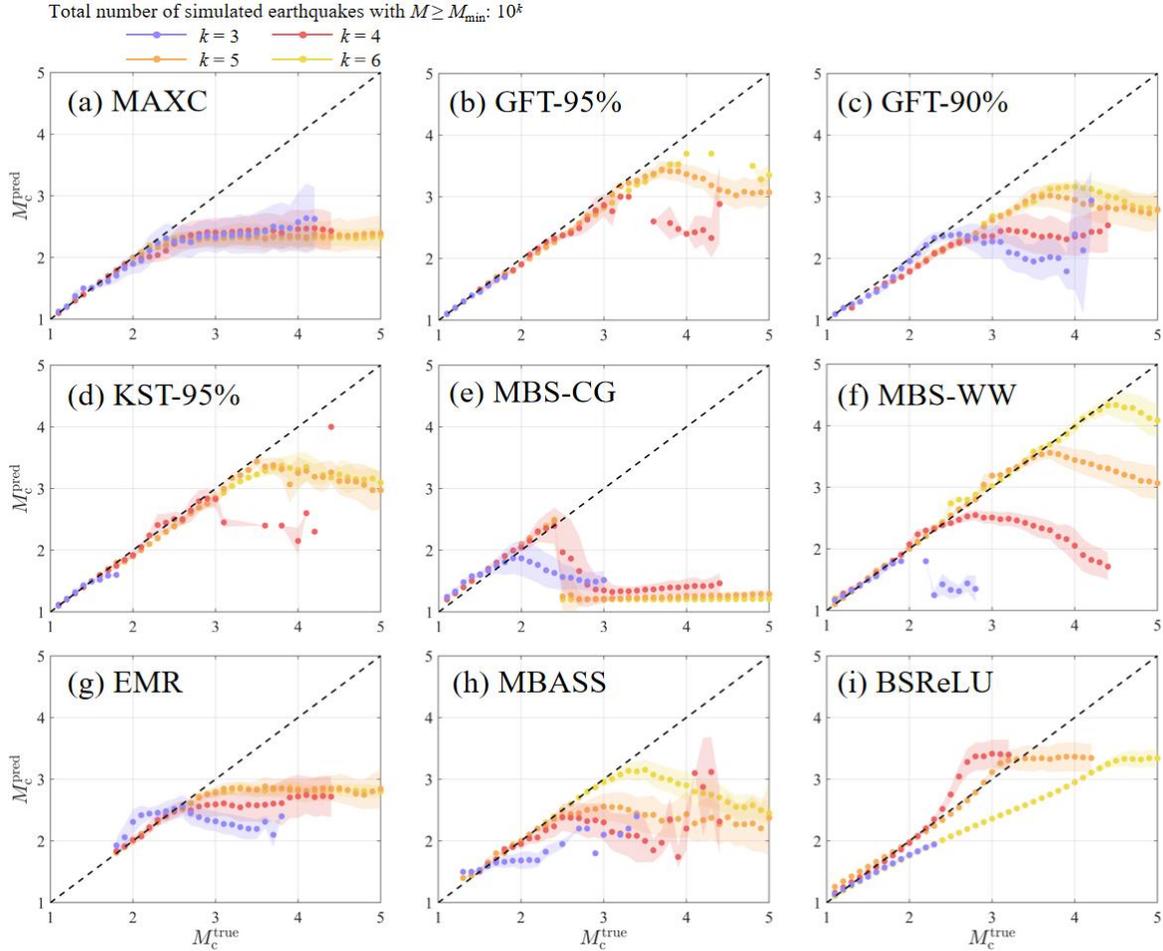

**Figure 2**. Results of the estimation of the completeness magnitude $M_c^{pred}$ as a function of the true completeness magnitude $M_c^{true}$ for the simulated datasets with homogeneous $M_c$, after being iteratively pruned three times, as shown in Figure 1 and Panels (IV) of Figure S1. Results are obtained using the following methods: (a) maximum curvature (MAXC) method, (b) goodness-of-fit test at 95% confidence level (GFT-95%), (c) goodness-of-fit test at 90% confidence level (GFT-90%), (d) Kolmogorov-Smirnov statistic test at 95% confidence level (KST-95%), (e) b-value stability method by Cao and Gao (2002) (MBS-CG), (f) $b$-value stability method by Woessner and Wiemer (2005) (MBS-WW), (g) entire magnitude range (EMR) method, (h) median-based analysis of the segment slope (MBASS), and (i) Rectified Linear Unit with Black-Scholes model (BSReLU). The shaded areas represent the uncertainties in the estimated completeness magnitude $M_c^{pred}$. The different colors correspond to the total number of simulated earthquakes with $M \geq M_{min} = 1$, with $10^k$ for $k = 3, 4, 5, 6$, same as in Figure 1.

4.2 Estimating $M_c$ for simulated seismicity with heterogeneous $M_c$

Figure 3 shows the results of the estimation of $M_c^{pred}$ as a function of the number of grids ($2^N$ for $N = 1, 2, ..., 10$) used to distribute the simulated earthquakes. Our pruning approach, which uses grids to simulate heterogeneity in $M_c$, essentially captures both the spatial and/or temporal variability of seismicity's $M_c$. Overall, except for MBS-WW, the other eight methods tend to yield very small $M_c^{pred}$ values when dealing with catalogs with heterogeneous $M_c$ distributions. These estimates are often far below the mean $M_c^{true}$ over all grids. Recall that our construction of the synthetic catalogs ensures that the mean $M_c^{true}$ over all grids is equal to 3.0, with half of the grids having $M_c^{true} \leq 3.0$, and the other half having $M_c^{true} > 3.0$. Only MBS-WW, when the dataset reaches a sufficient size (such as $k = 5$ and 6) and for higher-resolution grid scenarios (such as $2^N$ for $N \geq 3$), retrives $M_c^{pred}$ values that are close to the mean $M_c^{true}$ across all grids (i.e., 3.0). For all methods except MBS-WW and BSReLU, Mcpred, regardless of $N$, remain consistently between approximately 1.0 and 2.0. BSReLU, on the other hand, produces highly unstable $M_c^{pred}$, with significant fluctuations between 0.5 and 2.5.

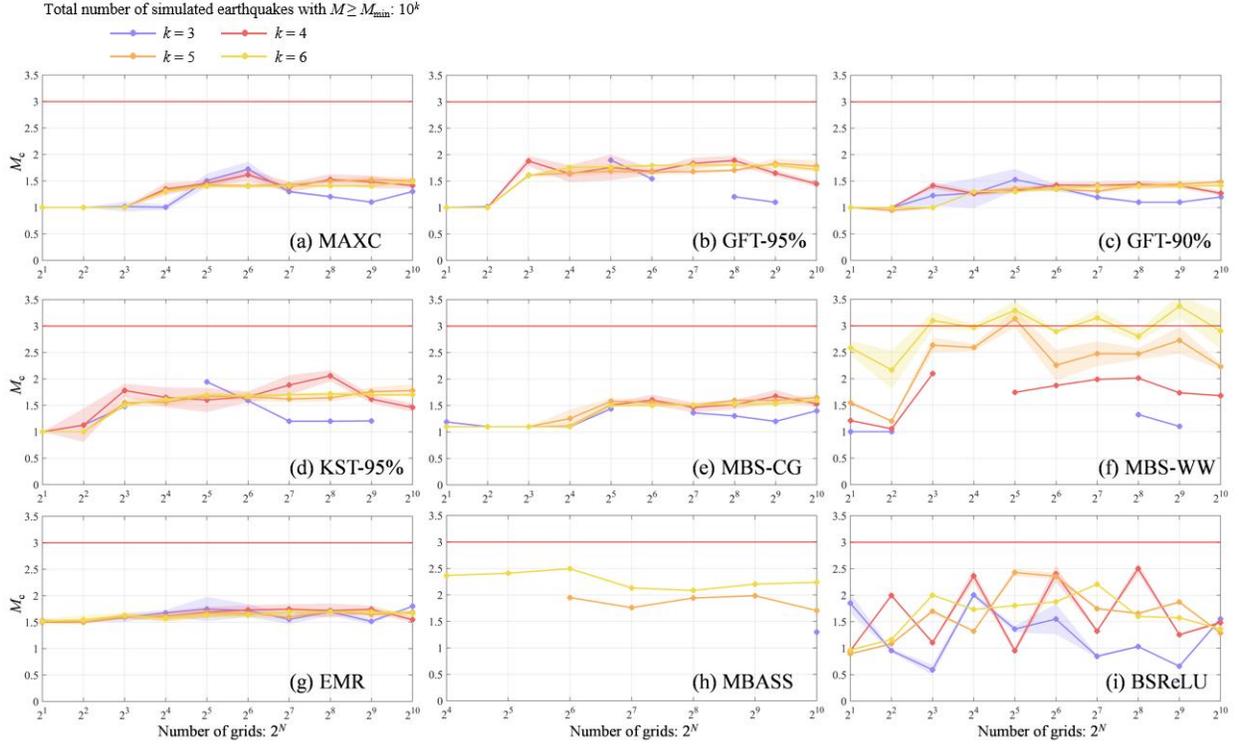

**Figure 3**. Results of the estimation of the completeness magnitude $M_c$ as a function of the number of grids used to distribute the simulated earthquakes. The total number of simulated earthquakes with $M \geq M_{min}$, given by $10^k$ for $k = 3, 4, 5, 6$, is randomly and uniformly distributed across $2^N$ ($N = 1, 2, ..., 10$) grids. The minimum completeness magnitude ($M_c^{min} = 1.0$) and maximum completeness magnitude ($M_c^{max} = 5.0$) are assigned to the first and last grids, respectively, with the remaining grids having completeness magnitudes linearly interpolated between these values. For example, with $N = 2^2$ grids, the completeness magnitudes are $M_c^{grid1} = 1.0$, $M_c^{grid2} = 2.3$, $M_c^{grid3} = 3.7$, and $M_c^{grid4} = 5.0$, ensuring that the average $M_c$ across all grids is

3.0 (indicated by the red horizontal line in the figure). The shaded areas represent the uncertainties in the estimated completeness magnitude. The different colors correspond to the total number of simulated earthquakes. For methods used, refer to Figure 2.

4.3 Estimating $M_c$ for empirical seismicity

For the six study regions with unknown $M_c^{true}$ and spatiotemporally heterogeneous $M_c$ in the observed earthquake catalogs, we used the nine catalog-based methods described above to estimate the mean $M_c^{pred}$ and its uncertainty for each region, derived from 200 bootstrapping iterations (Efron, 1979). Table 1 gives for eight methods the $M_c$ estimate and its associated uncertainty. For BSReLU, Table 1 additionally provides estimates for both $M_c$ and $\sigma M_c$, along with their uncertainties.

For the four study regions in the Chinese mainland (Beijing-Tianjin-Hebei region, the southeastern coastal region, the Sichuan-Yunnan region, and northern Xinjiang) and the two additional regions, California and New Zealand, the lowest mean $M_c^{pred}$ is given by MAXC (Beijing-Tianjin-Hebei and southeastern coastal), MBS-CG (Sichuan-Yunnan, northern Xinjiang and New Zealand), and GFT-90% (California). Conversely, the largest mean $M_c^{pred}$ values are consistently given by MBS-WW. For Beijing-Tianjin-Hebei, southeastern coastal, Sichuan-Yunnan, northern Xinjiang, and California, the mean $M_c^{pred}$ values from MBS-WW are 1.75, 1.44, 1.71, 1.87, and 1.84, respectively—these values are the closest to 1.8 among all the methods. Recall that the spatial boundaries of these regions were delineated using the $M_c \approx 1.8$ contour lines derived from the posterior $M_c$ map generated by the BMC method in Li et al. (2024), which applied Bayesian corrections to $M_c$ estimates using prior information about network density. For New Zealand, the largest mean $M_c^{pred}$, 3.13, is also provided by MBS-WW.

However, the standard deviation of $M_c^{pred}$ estimated by MBS-WW is relatively large compared to the other methods across the six study regions. The standard deviations for MBS-WW in Beijing-Tianjin-Hebei, southeastern coastal, Sichuan-Yunnan, northern Xinjiang, California, and New Zealand are 0.17 (third largest), 0.20 (third largest), 0.19 (largest), 0.32 (largest), 0.10 (third largest), and 0.35 (second largest), respectively. For the six study regions, the method with the smallest $M_c^{pred}$ standard deviation is KST-95% for Beijing-Tianjin-Hebei, EMR for southeastern coastal, MAXC for Sichuan-Yunnan, MBS-CG for northern Xinjiang, MAXC for California, and GFT-95% for New Zealand. Conversely, the method with the largest $M_c^{pred}$ standard deviation is MBASS for Beijing-Tianjin-Hebei and southeastern coastal, MBS-WW for Sichuan-Yunnan and northern Xinjiang, and BSReLU for California and New Zealand.

**Table 1.** Estimation results of the completeness magnitude $M_c$ and their associated uncertainties derived from 200 bootstrapping iterations, using nine methods across the six study regions.

| Region | MAXC | GFT-95% | GFT-90% | KST-95% | MBS-CG | MBS-WW | EMR | MBASS | BSReLU | |
|---|---|---|---|---|---|---|---|---|---|---|
| | | | | | | | | | $M_c$ | $\sigma_{Mc}$ |
| Beijing-Tianjin-Hebei | 0.49 ± 0.10 | 1.01 ± 0.03 | 0.69 ± 0.03 | 1.00 ± 0.01 | 0.86 ± 0.16 | 1.75 ± 0.17 | 0.80 ± 0.08 | 1.32 ± 0.64 | 1.25 ± 0.28 | 1.06 ± 0.07 |
| Southeastern coastal | 0.26 ± 0.06 | 0.92 ± 0.05 | 0.32 ± 0.04 | 0.88 ± 0.05 | 0.34 ± 0.06 | 1.44 ± 0.20 | 0.60 ± 0.01 | 1.35 ± 0.63 | 0.76 ± 0.34 | 1.00 ± 0.15 |
| Sichuan-Yunnan | 1.10 ± 0.00 | 1.30 ± 0.01 | 1.00 ± 0.00 | 1.30 ± 0.01 | 0.20 ± 0.00 | 1.71 ± 0.19 | 1.30 ± 0.00 | 1.60 ± 0.08 | 0.79 ± 0.13 | 0.56 ± 0.11 |
| Northern Xinjiang | 1.36 ± 0.06 | 1.59 ± 0.03 | 1.38 ± 0.04 | 1.60 ± 0.05 | 0.20 ± 0.00 | 1.87 ± 0.32 | 1.57 ± 0.08 | 1.70 ± 0.07 | 1.19 ± 0.28 | 0.52 ± 0.16 |
| California | 0.90 ± 0.00 | 0.91 ± 0.00 | 0.60 ± 0.00 | 0.91 ± 0.06 | 0.70 ± 0.00 | 1.84 ± 0.10 | 0.96 ± 0.10 | 0.91 ± 0.09 | 1.22 ± 0.49 | 0.89 ± 0.14 |
| New Zealand | 1.94 ± 0.05 | 2.17 ± 0.00 | 1.90 ± 0.00 | 2.10 ± 0.00 | 0.20 ± 0.00 | 3.13 ± 0.35 | 2.00 ± 0.05 | 2.17 ± 0.05 | 2.37 ± 0.44 | 0.72 ± 0.12 |

## 5 Discussion

In this section, we update the $M_c$ map from Li et al. (2024) using MBS-WW and further discuss the potential applications of BSReLU.

### 5.1 Estimating $M_c$ for empirical seismicity

Using the best-performing prior model, $M_c^{pred} = p_1 d(3)^{p_3} + p_2$, where $d(3)$ represents the distance to the third closest station and $p_1$, $p_2$, and $p_3$ are constants calibrated on the Chinese earthquake catalog from 2009 to 2022, Li et al. (2024) generated the spatial distribution of $M_c$ using the BMC method. The BMC method, proposed by Mignan et al. (2011), integrates observed $M_c^{obs}$ with prior Bayesian information derived from the relationship between seismic network density and $M_c$ observations, weighted by their respective standard deviations. Our results (e.g., Figures 2 and 3; Table 1) show that MBS-WW provides more accurate and reliable $M_c^{pred}$ compared to other methods. Consequently, we replaced MBASS used by Li et al. (2024) with MBS-WW. In this updated method, the calibrated constants $p_1$, $p_2$, and $p_3$ are 0.1264, 0.7022, and 0.4286, respectively, and the $M_c$ map for China was updated accordingly. Figure 4 shows: (a) the map of observed $M_c^{obs}$ obtained using MBS-WW for the period from 2009 to 2024, (b) the map of predicted $M_c^{pred}$ derived from the prior model, (c) the map of posterior $M_c^{post}$ estimated using the BMC method, and (d) the map of posterior standard deviations ($\sigma M_c^{post}$) estimated using the BMC method. We only included pixels where the number of events with $M \geq M_c^{pred}$ estimated by MBS-WW exceeded 300. This filtering criterion explains why many areas in Figure 4a do not display $M_c$ values.

The $M_c^{obs}$ map obtained using MBS-WW contains significantly fewer data points compared to the $M_c^{obs}$ map obtained by Li et al. (2024) using MBASS. This reduction in data volume suggests that many of the spatial data points for $M_c^{obs}$ derived from MBASS in Li et al. (2024) may be unreliable. MAXC is a standard method for estimating $M_c^{obs}$ in the BMC method (Mignan et al., 2011; Feng et al., 2022). However, it results in large uncertainty when applied to the Chinese catalog, with $\sigma M_c^{pred} = 0.56$, as reported by Mignan et al. (2013). The reason Li et al. (2024) used MBASS is that they found MBASS reduced these uncertainties compared to MAXC in the context of their study, with $\sigma M_c^{pred} = 0.47$, as reported by them. Our results show that the $\sigma M_c^{pred}$ obtained using MBS-WW is 0.43, which is a further improvement compared to the 0.47 reported by Li et al. (2024).

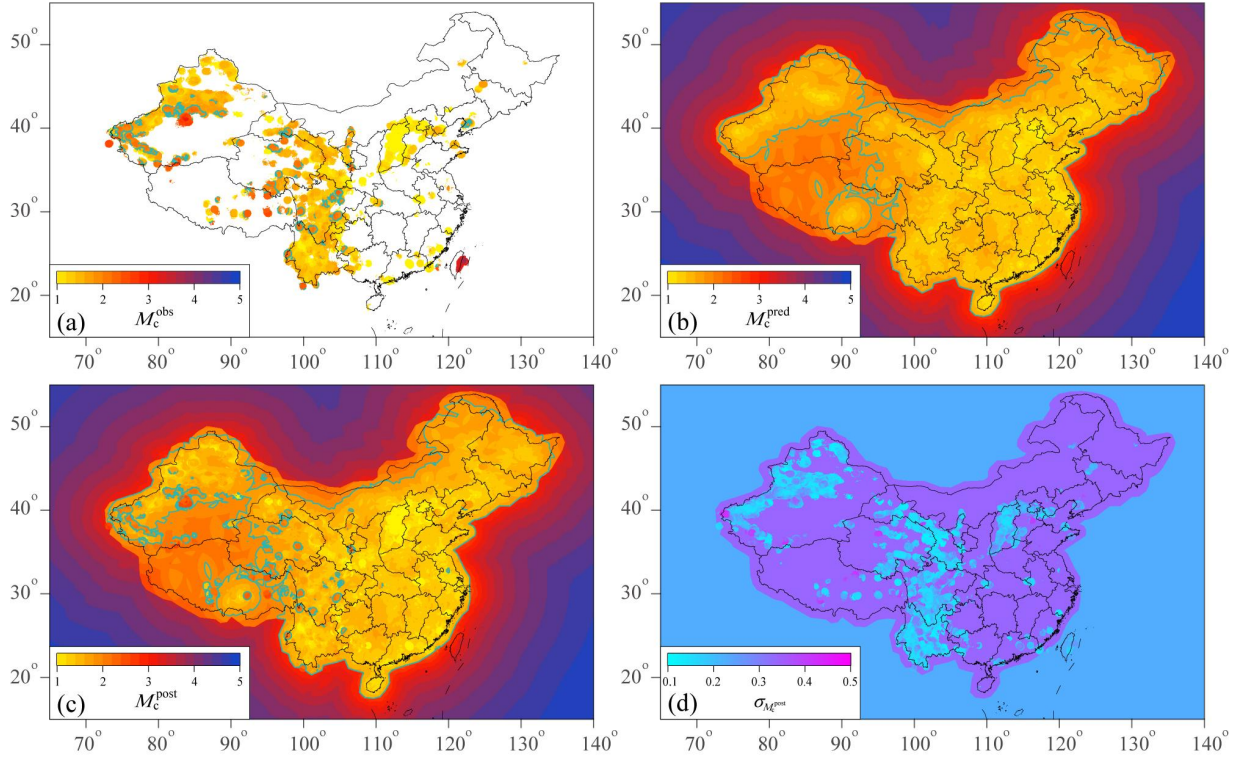

**Figure 4**. (a) Map of the observed $M_c^{obs}$ for the time period from 2009 to 2024. The estimated $M_c^{obs}$ of each grid is computed using the *b*-value stability method by Woessner and Wiemer (2005) (MBS-WW). (b) Map of the predicted $M_c^{pred}$ based on prior information for the existing broadband seismic stations. The standard deviations $\sigma M_c^{pred}$ of $M_c^{pred}$ inside and outside the Chinese mainland are 0.43 and 0.24, respectively, representing a reduction compared to ($\sigma M_c^{pred}$ = 0.47) reported by Li et al. (2024) using the median-based analysis of the segment slope (MBASS). (c) Map of the posterior $M_c^{post}$ estimated by the Bayesian Magnitude of Completeness (BMC) method for the time period from 2009 to 2022. (d) Map of the posterior standard deviations ($\sigma M_c^{post}$) estimated by the BMC method for the time period from 2009 to 2022.

5.2 BSReLU: From pointwise $M_c$ to distributed $M_c$

Within our evaluation framework, while not always the best-performing method, BSReLU delivers an impressive performance. For simulated datasets with homogeneous $M_c$, BSReLU demonstrates strong stability in its estimations of $M_c^{pred}$. However, deviations begin to appear starting around $M_c^{true} \approx 2.0$ for $k = 3$ and $k = 6$, and at $M_c^{true} = 2.3$ and 3.0 for $k = 4$ and $k = 5$, respectively. For simulated datasets with heterogeneous $M_c$, BSReLU produces highly unstable $M_c^{pred}$, exhibiting large fluctuations between 0.5 and 2.5. For empirical datasets, BSReLU often produces $M_c^{pred}$ that are relatively close to those of MBS-WW. It is important to highlight, however, that the evaluation framework used in this study is based on the assumption that the "cut-off $M_c$" concept is valid and fundamental. Specifically, this assumption posits that, above a certain $M_c$ value, data is 100% complete, while below this $M_c$ value, data is incomplete. Based on this assumption, we constructed the pruning function to generate datasets with a cut-off $M_c$. In contrast, the concept of $M_c$ in BSReLU is represented as a probability density function, specifically a Gaussian distribution defined by a mean $M_c$ and a standard deviation $\sigma M_c$. Therefore, our evaluation framework, which assumes a cut-off $M_c$, might be inherently unfair to BSReLU. In other words, the testing framework represents a mis-specification for BSReLU, while it is a correct specification for the other eight methods, as they also operate on the basis of the cut-off $M_c$ concept. When viewed in this context, it is particularly interesting to reassess the performance of BSReLU in this study.

Although EMR also incorporates a Gaussian complementary cumulative distribution function, it models the event detection probability of the seismic network rather than the completeness magnitude (Woessner and Wiemer, 2005). Specifically, EMR assumes that the probability of event detection by the network approaches 1 as $M$ is increased, and approaches 0 as $M$ is decreased. However, because the EMR framework cannot fully integrate the incomplete portion of FMD with the complete portion governed by the Gutenberg-Richter relationship, it still relies on the concept of a cut-off $M_c$. This assumes a step function where the event detection probability becomes exactly 1 for $M \geq M_c$, using Equation (1) to describe the CCFMD. In contrast, the BSReLU model overcomes this limitation. It simultaneously captures the Gutenberg-Richter behavior of the logarithm of the CCFMD, which asymptotically aligns with a linear trend for $M \geq M_c$, while also incorporating a probabilistic representation of the completeness magnitude. This allows the BSReLU model to account for the fact that, even for magnitudes larger than $M_c$, there is a small but nonzero probability that some large earthquakes might go undetected. Similarly, for magnitudes smaller than $M_c$, the probability of detecting smaller earthquakes is not zero but progressively diminishes as the magnitude decreases.

By integrating these features, the BSReLU model provides a more physically grounded framework for capturing the complexities of real-world earthquake catalogs, particularly those with incomplete entries. It offers greater flexibility and comprehensiveness in modeling compared to traditional approaches. Moreover, even for those seeking a straightforward and practical cut-off $M_c$ value, BSReLU can provide useful recommendations. For instance, based on the desired level of catalog completeness, one could adopt $M_c + \sigma M_c$, $M_c + 2\sigma M_c$, or $M_c + 3\sigma M_c$ to correspond to completeness probabilities of approximately 84%, 98%, and nearly 100%, respectively. This model holds significant potential for future applications, particularly in statistical seismology frameworks that require simultaneous consideration of the Gutenberg-Richter law and magnitude incompleteness. For example, it could be highly beneficial in models like the Epidemic-Type Aftershock Sequence (ETAS) model (Hawkes and Oakes, 1974; Kagan, 1991; Li et al., 2024; Ogata, 1988, 1998; Ogata and Zhuang, 2006; Sornette and Werner, 2005), which heavily depend on accurate representations of both complete and incomplete seismicity.

## 6 Conclusions

Analyzing the magnitude of completeness ($M_c$) is an indispensable first step in nearly all earthquake statistical analyses. In this process, catalog-based methods for estimating $M_c$ are the simplest, most straightforward, and most commonly used approach. Moreover, they often serve as the foundation for more advanced $M_c$ estimation techniques. Therefore, it is essential to identify which methods are most reliable and to determine the conditions under which they perform best. In this study, we evaluated and compared nine catalog-based methods, including two newly developed methods for estimating $M_c$, which leverage our expertise in statistical testing and computer science. These methods were systematically analyzed within a robust framework designed to evaluate their performance under varying conditions, including datasets with homogeneous and heterogeneous $M_c$ distributions, as well as real-world earthquake catalogs from regions such as China, California, and New Zealand.

For simulated earthquake catalogs with homogeneous $M_c$, all methods provide reasonably accurate $M_c^{pred}$ when $M_c^{true}$ is very low ($M_c^{true} \leq 2.0$). However, when $M_c^{true}$ is greater than or equal to 3.5, eight of the methods fail to produce reliable estimates, with the exception of MBS-WW. MBS-WW is the only method capable of generating reliable results when the minimum number of events in the complete portion is approximately 300 or more. For catalogs with heterogeneous $M_c$ distributions, the other eight methods tend to yield very small $M_c^{pred}$ values, which are often significantly below the mean $M_c^{true}$ across all grids. In contrast, MBS-WW achieves $M_c^{pred}$ estimates close to the mean $M_c^{true}$ (3.0) across all grids when datasets are sufficiently large (e.g., $k = 5$ and 6, where $k$ controls the number of simulated earthquake events given by $10^k$) and grid resolutions are high (e.g., $2^N$ for $N \geq 3$). For empirical earthquake catalogs across the six study regions, MBS-WW consistently produces the largest mean $M_c^{pred}$ values. In Beijing-Tianjin-Hebei, southeastern Coastal, Sichuan-Yunnan, northern Xinjiang, and California, the mean $M_c^{pred}$ values derived from MBS-WW are 1.75, 1.44, 1.71, 1.87, and 1.84, respectively. These estimates are the closest to the expected $M_c \approx 1.8$, which was used to delineate the spatial boundaries of these regions based on the posterior $M_c$ map generated by the BMC method in Li et al. (2024). However, the standard deviation of $M_c^{pred}$ estimates from MBS-WW is relatively larger compared to the other methods, ranging from 0.10 to 0.35 across the six regions. Based on these findings, we updated the $M_c$ map for China using MBS-WW. The previous map, generated by Li et al. (2024) using the MBASS method, was found to contain several potentially unreliable spatial data points due to MBASS's limitations in accurately capturing $M_c$. By replacing MBASS with MBS-WW, we reduced the uncertainties in $M_c$ estimates from 0.47 to 0.43. This improvement highlights the robustness of MBS-WW in generating more reliable and accurate spatial distributions of $M_c$.

We have introduced BSReLU as a new model to represent the technically and physically realistic progressive transition from the "complete" regime where all earthquakes are measured to the "incomplete" part of the catalogue where events are less and less likely to be detected, the smaller their magnitudes. Unlike traditional methods that rely on a strict cut-off $M_c$, BSReLU addresses the issue of incompleteness as a probabilistic concept. It models the likelihood of an earthquake being observed using a Gaussian distribution, defined by a mean $M_c$ and a standard deviation $\sigma M_c$. This approach offers significant advantages, especially in cases where the strict cut-off assumption oversimplifies the complexities of real-world seismicity. For instance, BSReLU captures the gradual transition between incomplete and complete data, reflecting the inherent uncertainties in event detection probabilities. Even for magnitudes larger than $M_c$, BSReLU accounts for the possibility of missing events, while for smaller magnitudes, it effectively models the decreasing probability of detection as the magnitude diminishes. This probabilistic framework provides a more realistic representation of seismicity, particularly for catalogs with varying levels of completeness. Unlike traditional methods, including MBS-WW, which assume that all data above a certain $M_c$ are 100% complete, BSReLU challenges this assumption by offering a more flexible and probabilistic perspective on completeness. This flexibility enables BSReLU to bridge the gap between traditional cut-off models and probabilistic frameworks, making it particularly well-suited for applications requiring a detailed understanding of data completeness. Note that BSReLU is just one implementation of the concept that the transition from complete to incomplete ought to be described by a gradual probabilistic model. Many parameterisations are possible, for instance by replacing the cumulative Gaussian function by other sigmoid functions with different left and right tails. Given these promising attributes, we believe that continued attention and exploration of this class of models are warranted for future research and applications. Moreover, novel testing methods need to be developed that are adapted to their gradual probabilistic structure.

**Acknowledgments**

The authors would like to thank Yang Zang from the China Earthquake Networks Center for his assistance in accessing the Chinese earthquake catalog data, as well as Prof. Zhongliang Wu, Prof. Changsheng Jiang, and Prof. Jiancang Zhuang for their valuable suggestions. This work is partially supported by the Guangdong Basic and Applied Basic Research Foundation (Grant No. 2024A1515011568), Shenzhen Science and Technology Innovation Commission (Grant no. GJHZ20210705141805017), and the Center for Computational Science and Engineering at Southern University of Science and Technology.

**Open Research**

The California catalog is sourced from the Advanced National Seismic System (ANSS) Comprehensive Earthquake Catalog (ComCat), accessible at https://earthquake.usgs.gov/data/comcat/ (last accessed: December 18, 2023). For the catalog in New Zealand, the data are obtained from the GeoNet Earthquake Catalog of New Zealand, accessible at https://quakesearch.geonet.org.nz/ (last accessed: December 18, 2023). We acknowledge the New Zealand GeoNet programme and its sponsors EQC, GNS Science, LINZ, NEMA and MBIE for providing data used in this study. The data for the China Seismic Experimental Site are acquired from the China Earthquake Networks Center (CENC) through the internal link provided by the Earthquake Cataloging System at China Earthquake Administration, available at http://10.5.160.18/console/index.action (last accessed: December 18, 2023), with a

Digital Object Identifier (DOI) of 10.11998/SeisDmc/SN. This data is not publicly available; it can be requested from the CENC.

**References**

Amorèse, D. (2007), Applying a Change-Point Detection Method on Frequency-Magnitude Distributions. *Bulletin of the Seismological Society of America*, 97(5), 1742–1749. doi:10.1785/0120060181

Arshad, M., Rasool, M. T., & Ahmad, M. I. (2002), Kolmogorov Smirnov Test for Generalized Pareto Distribution. *Journal of Applied Sciences*, 2, 488–490. doi:10.3923/jas.2002.488.490

Baumgartner, D., & Kolassa, J. (2021), Power considerations for Kolmogorov–Smirnov and Anderson–Darling two-sample tests. *Communications in Statistics - Simulation and Computation*, 52(7), 3137–3145. doi:10.1080/03610918.2021.1928193

Black, F., & Scholes, M. (1973), The pricing of options and corporate liabilities. *Journal of Political Economy*, 81(3), 637–654. doi:10.1086/260062

Cao, A. M., & Gao, S. S. (2002), Temporal variations of seismic b-values beneath northeastern Japan island arc. Geophysical Research Letters., 29(9), 48–1–48–3. doi:10.1029/2001GL013775

Clauset, A., Shalizi, C. R., & Newman, M. E. J. (2009), Power-law distributions in empirical data. *SIAM Review*, 51(4), 661–703. doi:10.1137/070710111

Corral, A., Font, F., & Camacho, J. (2011), Non-characteristic half-lives in radioactive decay. *Phys Rev E Stat Nonlin Soft Matter Phys.*, 83(6 Pt 2), 066103. doi:10.1103/PhysRevE.83.066103

Efron, B. (1979), Bootstrap Methods: Another Look at the Jackknife. Ann. Statist., 7(1), 1–26.

García-Hernández, R., D'Auria, L., Barrancos, J., & Padilla, G. D. (2019), On the functional expression of frequency-magnitude distributions: A comprehensive statistical examination. *Bulletin of the Seismological Society of America*, 109(1), 482–486. doi:10.1785/0120180197

Gibbons, J. D., & Chakraborty, S. (2010), Nonparametric statistical inference. In: Lovric, M. (eds) *International Encyclopedia of Statistical Science*. doi:10.1201/9781439896129

Godano, C., Convertito, V., & Pino, N. A. (2021), The signal-to-noise ratio and the completeness magnitude: The effect of the COVID-19 lockdown. *Seismological Research Letters*, 12(5), 525. doi:10.3390/atmos12050525

Hawkes, A. G., & Oakes, D. (1974), A cluster process representation of a self-exciting process. *Journal of Applied Probability*, 11(3), 493–503. doi:10.2307/3212693

Herrmann, M., & Marzocchi, W. (2021), Inconsistencies and lurking pitfalls in the magnitude–frequency distribution of high-resolution earthquake catalogs. *Seismological Research Letters*, 92(2), 909–922. doi:10.1785/0220200337

Iwata, T. (2008), Low detection capability of global earthquakes after the occurrence of large earthquakes: Investigation of the Harvard CMT catalogue. *Geophysical Journal International*, 174(3), 849–856. doi: 10.1111/j.1365-246X.2008.03864.x

Kagan, Y. Y. (1991b), Likelihood analysis of earthquake catalogues. *Geophysical Journal International*, 106(1), 135–148. doi:10.1111/j.1365-246X.1991.tb04607.x

Kagan, Y. Y. (2003), Accuracy of modern global earthquake catalogs. *Physics of the Earth and Planetary Interiors*, 135(2–3), 173–209. doi:10.1016/S0031-9201(02)00214-5.

Lanzante, J. R. (1996), Resistant, robust and non-parametric techniques for the analysis of climate data: Theory and examples, including applications to historical radiosonde station data. *International Journal of Climatology*, 16(11), 1197–1226.

https://doi.org/10.1002/(SICI)1097-0088(199611)16:11<1197::AID-JOC89>3.0.CO;2-L

Li et al. (2023), Predicting the future performance of the planned seismic network in Chinese mainland. *Seismological Research Letters*. 94 (6): 2698–2711. doi:10.1785/0220230102

Li, J., Sornette, D., Wu, Z., et al. (2024), Revisiting seismicity criticality: A new framework for bias correction of statistical seismology model calibrations. *Journal of Geophysical Research: Solid Earth* (under review, arXiv: https://arxiv.org/abs/2404.16374).

Lilliefors, H. W. (1969), On the Kolmogorov-Smirnov test for the exponential distribution with mean unknown. *Journal of the American Statistical Association*, 64(325), 387–389.

https://api.semanticscholar.org/CorpusID:121226377

Lombardi, A. M. (2021), A normalized distance test for co-determining the completeness magnitude and b-value of earthquake catalogs. *Journal of Geophysical Research: Solid Earth*, 126(3). doi:10.1029/2020JB021242

Mignan, A. (2012), Functional shape of the earthquake frequency-magnitude distribution and completeness magnitude. *Journal of Geophysical Research*, 117(B8). doi:10.1029/2012JB009347

Mignan, A., & Woessner, J. (2012), Estimating the magnitude of completeness for earthquake catalogs. *Community Online Resource for Statistical Seismicity Analysis*. doi:10.5078/corssa-00180805.

Molnár, L., & Szokol, P. (2014), Kolmogorov-Smirnov isometries of the space of generalized distribution functions. *Mathematica Slovaca*, 64(2), 433–444. doi:10.2478/s12175-014-0216-8

Ogata, Y. (1988), Statistical models for earthquake occurrences and residual analysis for point processes. *Journal of the American Statistical Association*, 83(401), 9–27. doi:10.1080/01621459.1988.10478560

Ogata, Y. (1998), Space-time point-process models for earthquake occurrences. *Annals of the Institute of Statistical Mathematics*, 50(2), 379–402. doi:10.1023/A:1003403601725

Ogata, Y., & Zhuang, J. (2006), Space-time ETAS models and an improved extension. *Tectonophysics*, 413(1–2), 13–23. doi:10.1016/j.tecto.2005.10.016.

Ogata, Y., & Katsura, K. (1993), Analysis of temporal and spatial heterogeneity of magnitude frequency distribution inferred from earthquake catalogues. *Geophysical Journal International*, 113, 727–738. doi:10.1111/j.1365-246X.1993.tb04663.x

Ogata, Y., & Katsura, K. (2006), Immediate and updated forecasting of aftershock hazard. *Geophysical Research Letters*, 33(10). doi:10.1029/2006GL025888

Pavlenko, V. A., & Zavyalov, A. D. (2022), Comparative Analysis of the Methods for Estimating the Magnitude of Completeness of Earthquake Catalogs. Izv., *Physics of the Solid Earth*, 58, 89–105. doi:10.1134/S1069351322010062

Petersen, T., Gledhill, K., Chadwick, M., Gale, N. H., & Ristau, J. (2011), The New Zealand National Seismograph Network. *Seismological Research Letters*, 82(1), 9–20. doi:10.1785/gssrl.82.1.9

Rydelek, P. A., & Sacks, I. S. (1989), Testing the completeness of earthquake catalogues and the hypothesis of self-similarity. *Nature*, 337(6204), 251–253. doi:10.1038/337251a0

Shi, Y., & Bolt, B. A. (1982), The standard error of the magnitude-frequency b-value. *Bulletin of the Seismological Society of America*, 72(5), 1677–1687. doi:10.1785/BSSA0720051677

Schorlemmer, D., & Woessner, J. (2008), Probability of detecting an earthquake. *Bulletin of the Seismological Society of America*, 98(5), 2103–2117. doi:10.1785/0120070105

Siegel, S., & Castellan, N. (1988), Nonparametric Statistics for the Behavioral Sciences. McGraw-Hill.

Sornette, D., & Werner, M. J. (2005), Apparent clustering and apparent background earthquakes biased by undetected seismicity. *Journal of Geophysical Research: Solid Earth*, 110(B9), B09303, doi:10.1029/2005JB00362. doi:10.1029/2005JB003621

Taroni, M. (2023), Estimating the magnitude of completeness of earthquake catalogs using a simple random variable transformation. *The Seismic Record*, 3(3), 194–199. doi:10.1785/0320230017

Tormann, T., Wiemer, S., & Mignan, A. (2014), Systematic survey of high-resolution *b* value imaging along Californian faults: Inference on asperities. *Journal of Geophysical Research: Solid Earth*, 119(3), 2029–2054. doi:10.1002/2013JB010867

U.S. Geological Survey (2017), Advanced National Seismic System (ANSS) Comprehensive Catalog of Earthquake Events and Products. Various. https://doi.org/10.5066/F7MS3QZH.GNS.

Wilcoxon, F. (1945), Individual comparisons by ranking methods. *Biometrics*, 1(8), 80–83.doi:https://api.semanticscholar.org/CorpusID:53662922

Wiemer, S. (2001), A software package to analyze seismicity: ZMAP. *Seismological Research Letters*, 72(3), 373–382. doi:10.1785/gssrl.72.3.373

Wiemer, S., & Wyss, M. (2000), Minimum Magnitude of Completeness in Earthquake Catalogs: Examples from Alaska, the Western United States, and Japan. *Bulletin of the Seismological Society of America*, 90(4), 859–869. doi:10.1785/0119990114

Wiemer, S., & Wyss, M. (2003), Reply to "Comment on 'Minimum magnitude of completeness in earthquake catalogs: examples from Alaska, the Western United States and Japan,' by Stefan Wiemer and Max Wyss," by Paul A. Rydelek and I. S. Sacks. *Bulletin of the Seismological Society of America*, 93(4), 1868–1871. doi:10.1785/0120020103

Wesnousky, S. G. (1994), The Gutenberg-Richter or Characteristic Earthquake Distribution, Which Is It? *Bulletin of the Seismological Society of America*, 84(6), 1940–1958. doi:10.1785/BSSA0840061940

Woessner, J., & Wiemer, S. (2005), Assessing the quality of earthquake catalogues: Estimating the magnitude of completeness and its uncertainty. *Bulletin of the Seismological Society of America*, 95(2), 684–698. doi:10.1785/0120040007

Wyss, M., Hasegawa, A., Wiemer, S., & Umino, N. (1999), Quantitative mapping of precursory seismic quiescence before the 1989, M 7.1 off-Sanriku earthquake, Japan. Annals of Geophysics, 42(5), 851–869. doi:10.4401/ag-3765

Supporting Information for

**Estimating Magnitude Completeness in Earthquake Catalogs:
A Comparative Study of Catalog-based Methods**

Xinyi Wang[1,2], Jiawei Li[2], Feng Ao[1], and Didier Sornette[2]

[1] School of Computer Science, Chengdu University of Information Technology (CUIT), Chengdu, China.
[2] Institute of Risk Analysis, Prediction and Management (Risks-X), Academy for Advanced Interdisciplinary Studies, Southern University of Science and Technology (SUSTech), Shenzhen, China.

**Content of this file**

Figures S1 to CCFMD and FMD alongside the Gutenberg-Richter (GR) law for different total number of simulated earthquakes.

Figures S2 to CCFMD and FMD for the observed seismicity.

Figures S3 to Results of $M_c^{\text{pred}}$ as a function of $M_c^{\text{true}}$ for the simulated datasets with homogeneous $M_c$, after being iteratively pruned one times.

Figures S4 to Results of $M_c^{\text{pred}}$ as a function of $M_c^{\text{true}}$ for the simulated datasets with homogeneous $M_c$, after being iteratively pruned two times.

Figures S5 to Results of $M_c^{\text{pred}}$ as a function of $M_c^{\text{true}}$ for the simulated datasets with homogeneous $M_c$, after being iteratively pruned four times.

## Introduction

In the present study, we assess nine catalog-based methods, including two newly developed approaches, under a robust evaluation framework specifically tailored for simulating datasets with non-uniform $M_c$ distributions across both spatial and temporal dimensions. The Supporting Information accompanying this study provides an in-depth statistical characterization of both simulating and observed seismicity, along with results of the calibrated $M_c^{pred}$ as a function of $M_c^{true}$ used in generating the simulated datasets with homogeneous $M_c$, after being iteratively pruned one, two and four times, respectively, as explained in the main text.

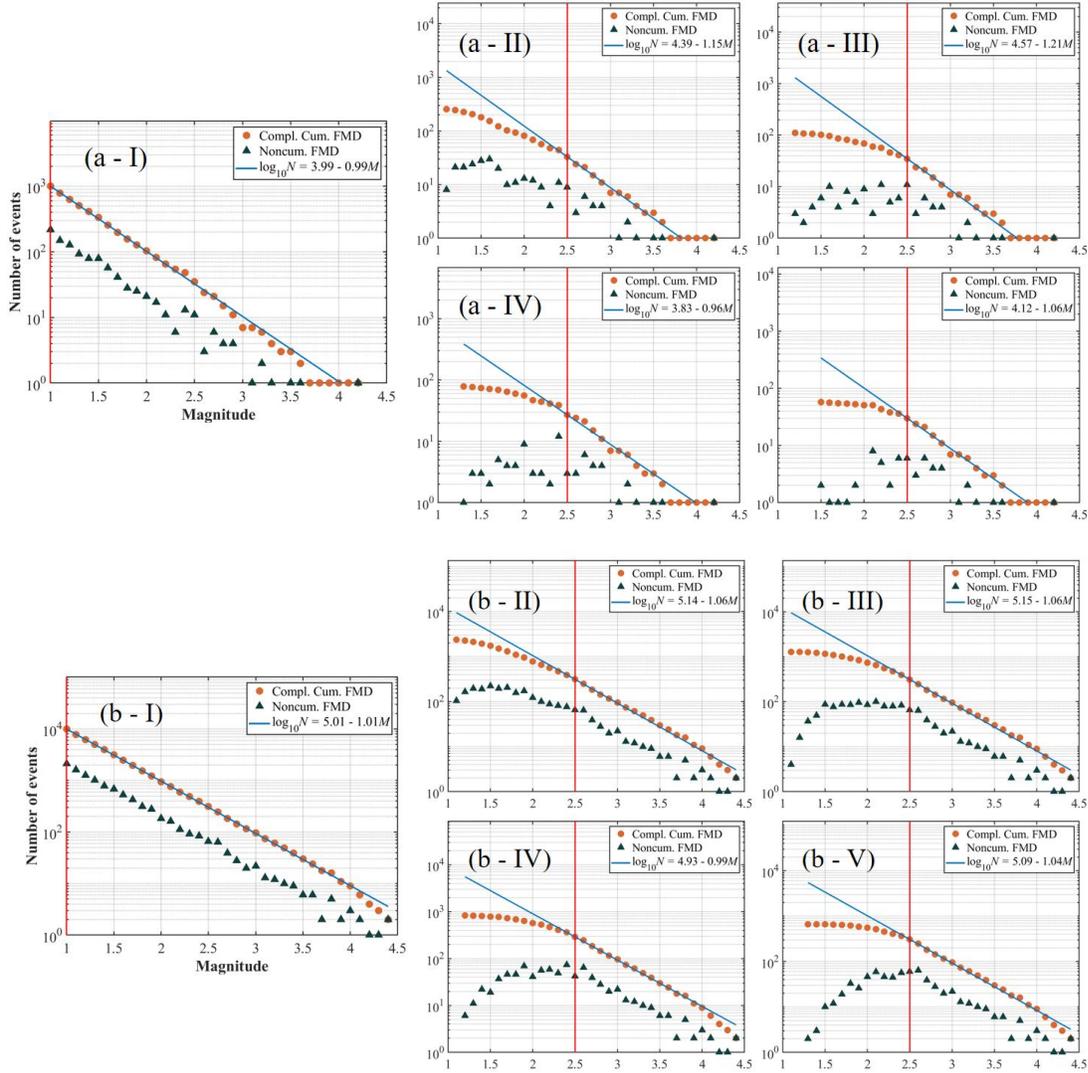

**Figure S1.** Complementary cumulative distribution (CCFMD) and density distribution (FMD) alongside the Gutenberg-Richter (GR) law, fitted using simulated earthquakes with magnitudes larger than 2.5, for different total number of simulated earthquakes with $M \geq M_{min} = 1.0$: (a) $10^3$, (b) $10^4$, (c) $10^5$, and (d) $10^6$. For each case, the $b$-value of the GR law is set to 1.0 to simulate synthetic seismicity, and panel (I) shows the CCFMD and FMD, which follow a pure power law for the entire magnitude range from $M_{min}$ to $M_{max}$. Panels (II) to (V) show results from iteratively applying the pruning function to magnitudes from $m_{min} = 1.0$ to $m_{max} = 2.5$, one to four times, respectively. By setting $M_c^{true}$ to $m_{max} = 2.5$, the data are separated into incomplete and power law (complete) parts. In this study, $m_{max}$ (i.e., $M_c^{true}$) is varied between 1.0 and 5.0, and each data set, like the one in Figure S1, is generated 200 times. For details on data pruning, refer to the main text.

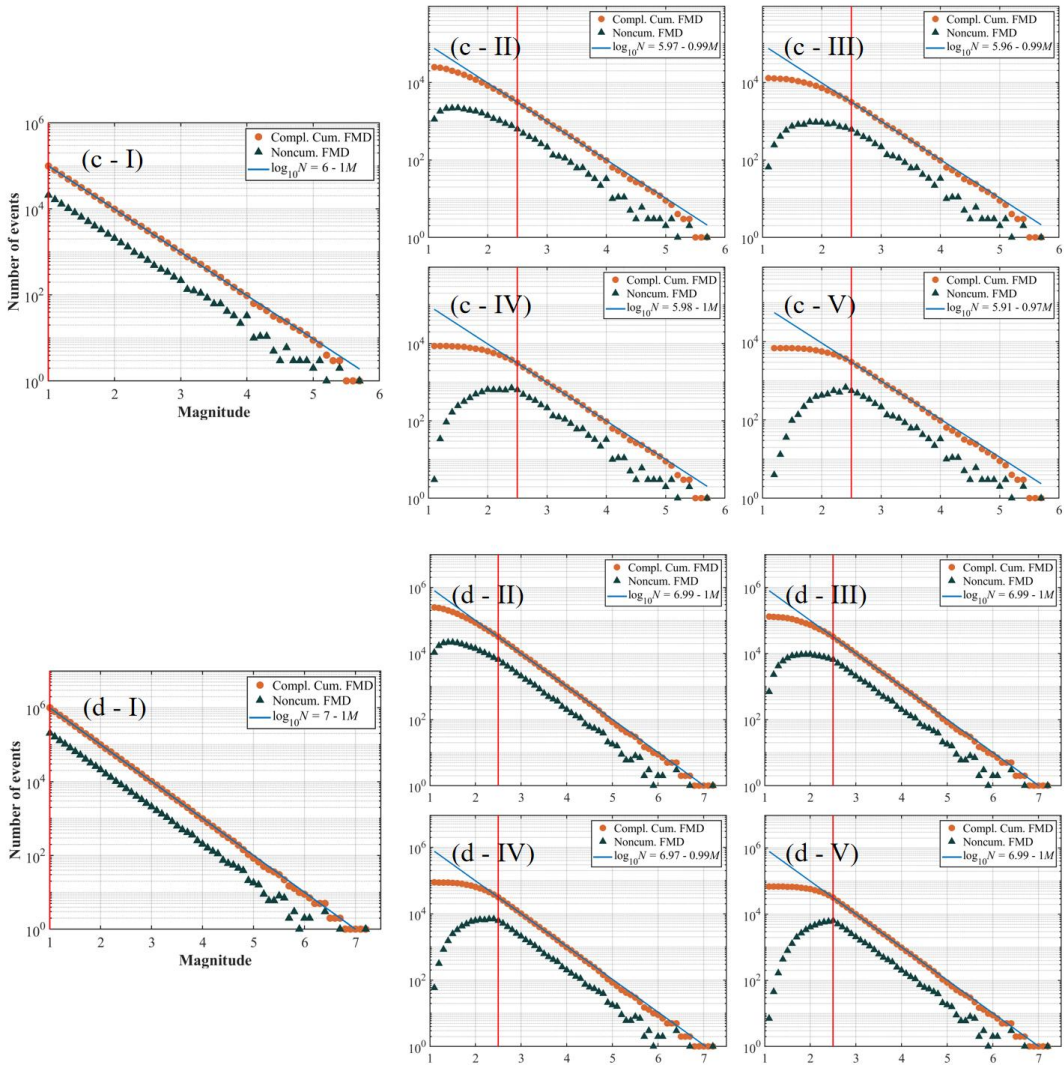

**Figure S1.** -continued.

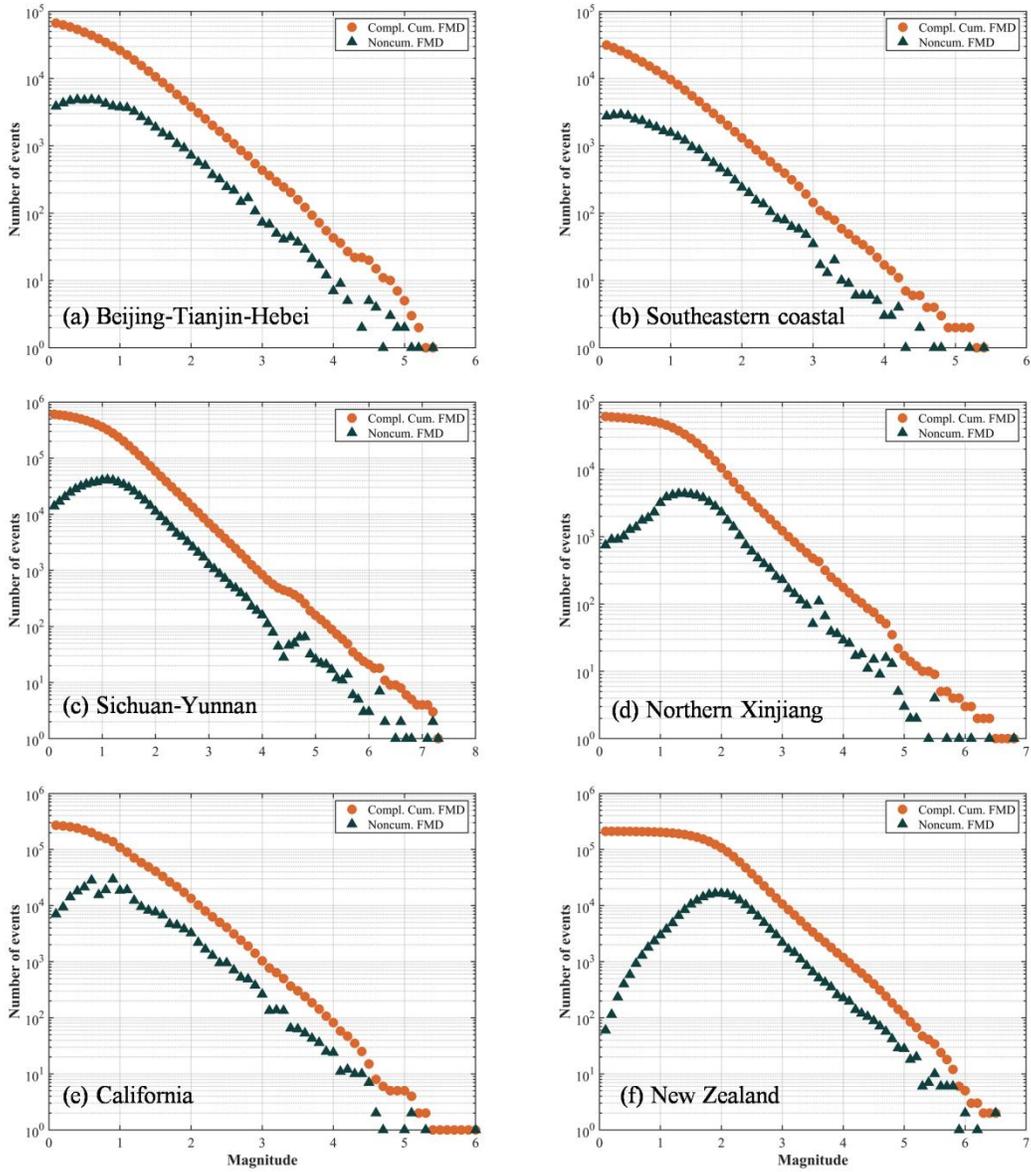

**Figure S2.** Complementary cumulative distribution (CCFMD) and density distribution (FMD) for the observed seismicity in (a) the Beijing-Tianjin-Hebei region, (b) the southeastern coastal region, (c) the Sichuan-Yunnan region, (d) northern Xinjiang, (e) California, and (f) New Zealand.

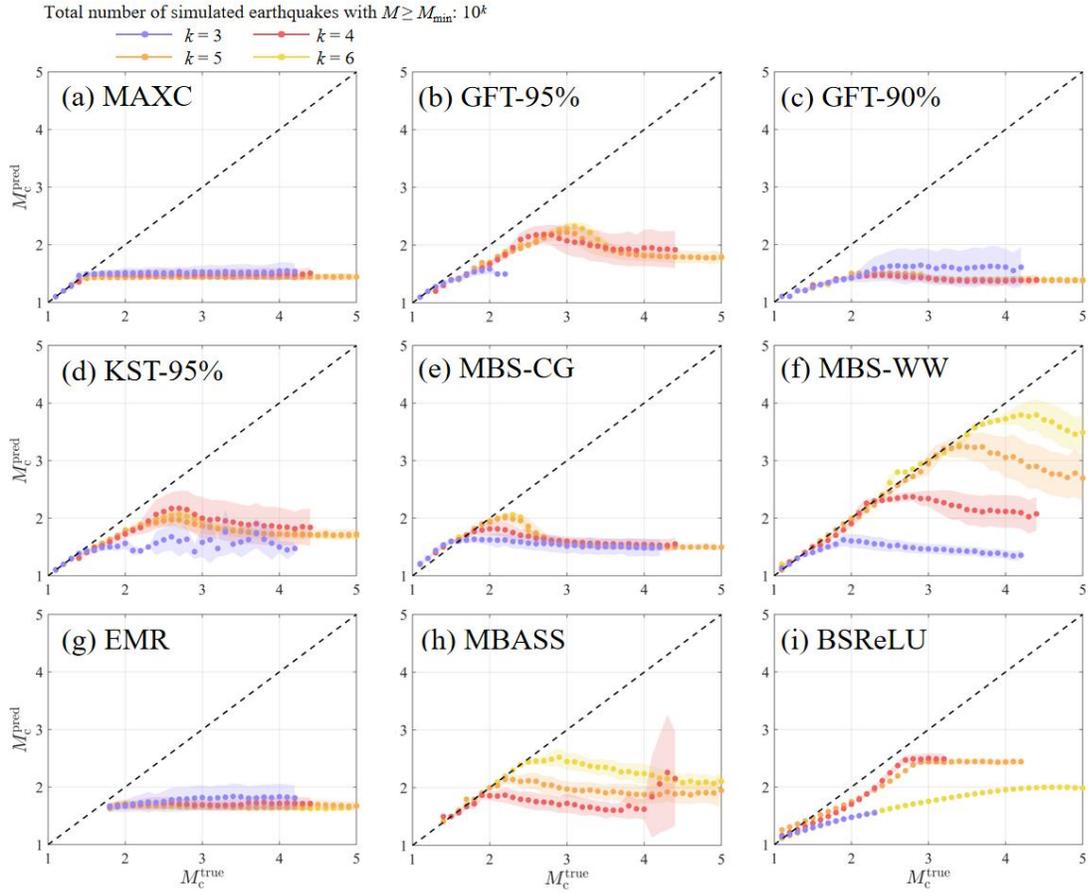

**Figure S3.** Results of the estimation of the completeness magnitude $M_c^{pred}$ as a function of the true completeness magnitude $M_c^{true}$ for the simulated datasets with homogeneous $M_c$, after being iteratively pruned one time, as shown in Panels (II) of Figure S1. Other details are the same as those in Figure 2.

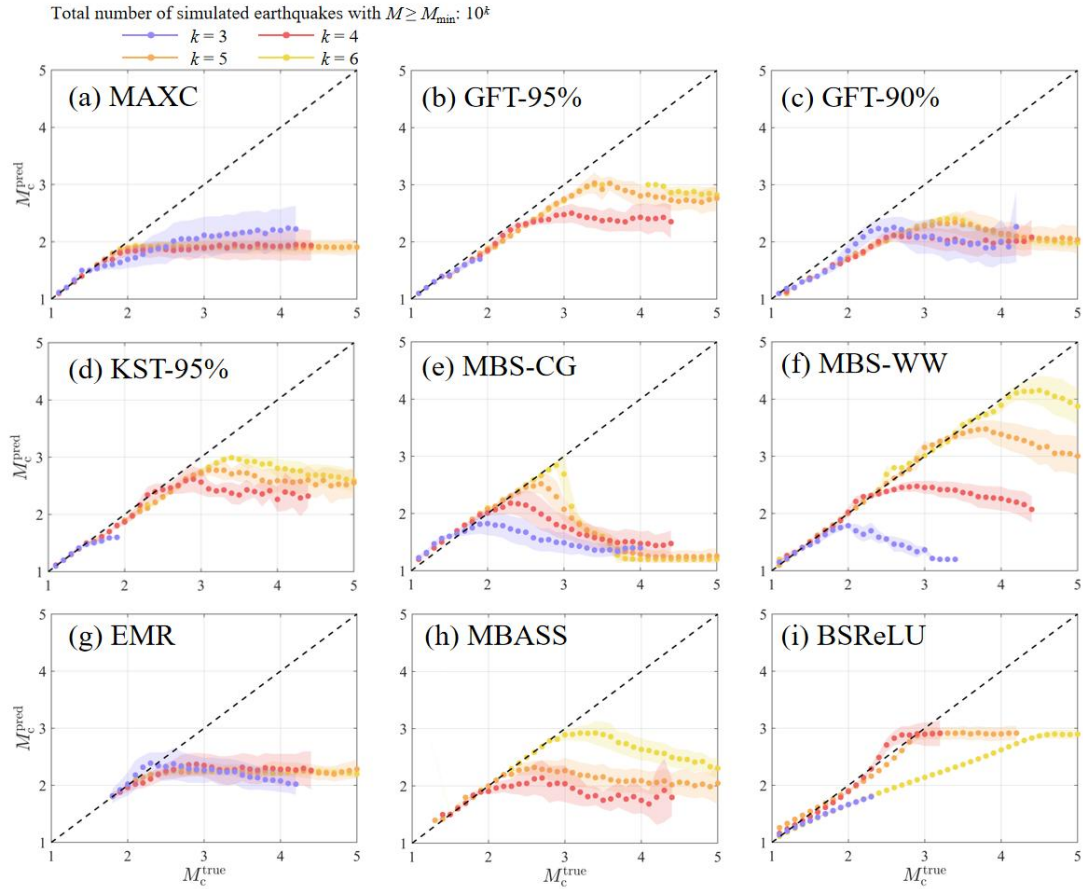

**Figure S4.** Results of the estimation of $M_c^{pred}$ as a function of $M_c^{true}$ for the simulated datasets with homogeneous $M_c$, after being iteratively pruned two times, as shown in Panels (III) of Figure S1. Other details are the same as those in Figure 2.

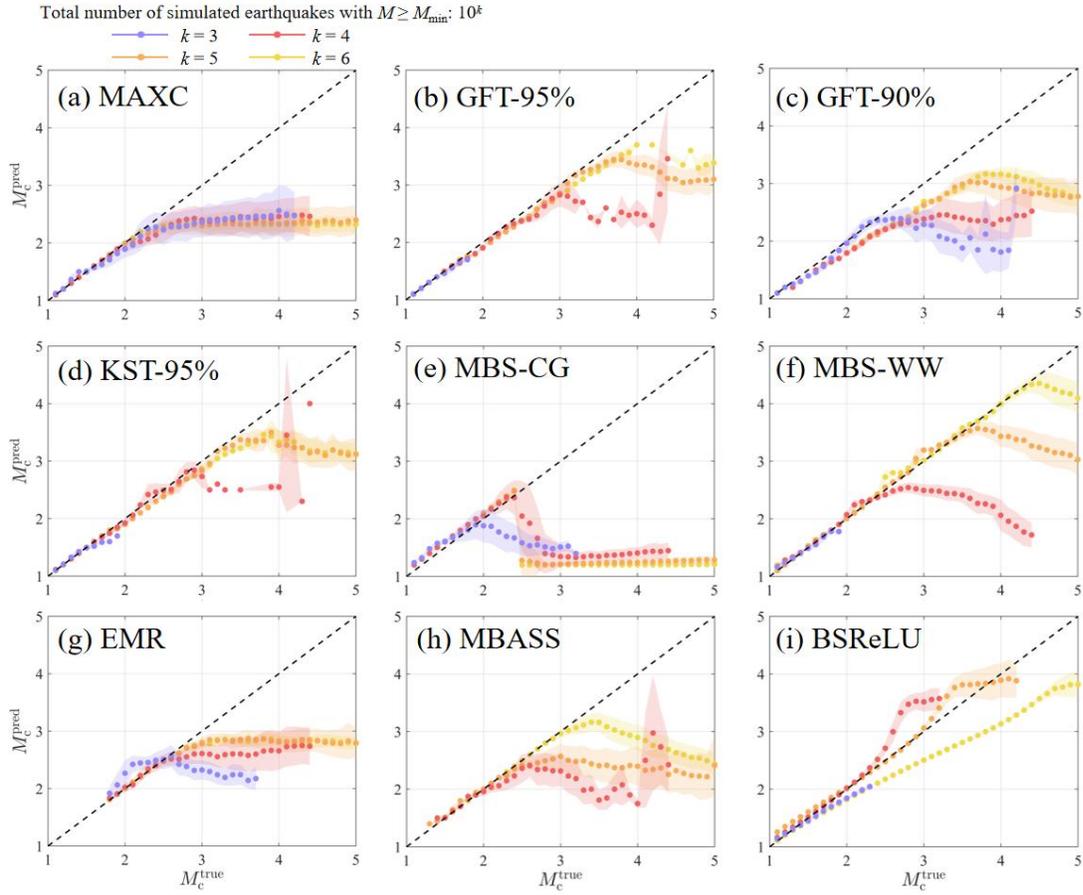

**Figure S5.** Results of the estimation of $M_c^{pred}$ as a function of $M_c^{true}$ for the simulated datasets with homogeneous $M_c$, after being iteratively pruned four times, as shown in Panels (V) of Figure S1. Other details are the same as those in Figure 2.